\documentclass[aps,pre,twocolumn,superscriptaddress,showkeys,showpacs]{revtex4}
\usepackage{graphicx, epsfig}
\usepackage{epstopdf}
\usepackage{amssymb, amsmath, amsfonts, multirow, rotate, color}
\usepackage{listings}
\usepackage{color}

\definecolor{mygreen}{rgb}{0,0.6,0}
\definecolor{mygray}{rgb}{0.5,0.5,0.5}
\definecolor{mymauve}{rgb}{0.58,0,0.82}

\DeclareGraphicsExtensions{.  jpg, .  pdf,  .  mps,  . png,  . eps,  .  ps,  .  EPS}
\usepackage{amsmath}
\usepackage{color}
\begin{document}
\newcommand{\avg}[1]{\langle{#1}\rangle}
\newcommand{\ket}[1]{\left |{#1}\right \rangle}
\newcommand{\beq}{\begin{equation}}
\newcommand{\eneq}{\end{equation}}
\newcommand{\beqnn}{\begin{equation*}}
\newcommand{\eneqnn}{\end{equation*}}
\newcommand{\beqy}{\begin{eqnarray}}
\newcommand{\eneqy}{\end{eqnarray}}
\newcommand{\beqynn}{\begin{eqnarray*}}
\newcommand{\eneqynn}{\end{eqnarray*}}
\newcommand{\half}{\mbox{$\textstyle \frac{1}{2}$}}
\newcommand{\proj}[1]{\ket{#1}\bra{#1}}
\newcommand{\av}[1]{\langle #1\rangle}
\newcommand{\braket}[2]{\langle #1 | #2\rangle}
\newcommand{\bra}[1]{\langle #1 | }
\newcommand{\Avg}[1]{\left\langle{#1}\right\rangle}
\newcommand{\inprod}[2]{\braket{#1}{#2}}
\newcommand{\upket}{\ket{\uparrow}}
\newcommand{\downket}{\ket{\downarrow}}
\newcommand{\Tr}{\mathrm{Tr}}
\newcommand{\hcontrol}{*!<0em, . 025em>-=-{\Diamond}}
\newcommand{\hctrl}[1]{\hcontrol \qwx[#1] \qw}
\newenvironment{proof}[1][Proof]{\noindent\textbf{#1. } }{\ \rule{0. 5em}{0. 5em}}
\newtheorem{mytheorem}{Theorem}
\newtheorem{mylemma}{Lemma}
\newtheorem{mycorollary}{Corollary}
\newtheorem{myproposition}{Proposition}
\newcommand{\vp}{\vec{p}}
\newcommand{\Or}{\mathcal{O}}
\newcommand{\so}[1]{{\ignore{#1}}}

\newcommand{\red}[1]{\textcolor{red}{#1}}
\newcommand{\blue}[1]{\textcolor{blue}{#1}}

\newcommand{\bea}{\begin{eqnarray}}
\newcommand{\eea}{\end{eqnarray}}
\newcommand{\gt}{\tilde{g}}
\newcommand{\mt}{\tilde{\mu}}
\newcommand{\et}{\tilde{\varepsilon}}
\newcommand{\ct}{\tilde{C}}
\newcommand{\bt}{\tilde{\beta}}

\title{Generalized network structures: \\
The configuration model  and the canonical ensemble of simplicial complexes}
\author{Owen T. Courtney}
\affiliation{School of Mathematical Sciences, Queen Mary University of London, E1 4NS, London, UK}
\author{Ginestra Bianconi}
\affiliation{School of Mathematical Sciences, Queen Mary University of London, E1 4NS, London, UK}

\begin{abstract}
Simplicial complexes are generalized network structures able to encode interactions occurring between more than two nodes. Simplicial complexes describe a large variety of complex interacting systems ranging from brain networks, to social and collaboration networks. Here we characterize the structure of simplicial complexes using their {\em generalized degrees} that capture fundamental properties of one, two, three  or more linked nodes. Moreover we introduce  the configuration model and the canonical ensemble of simplicial complexes, enforcing respectively the sequence of generalized degrees of the nodes and  the sequence of the expected generalized degrees of the nodes. We  evaluate  the entropy of these ensembles, finding the asymptotic expression for the number of simplicial complexes in the configuration model.  We provide the algorithms for the construction of simplicial complexes belonging to  the configuration model and the canonical  ensemble of simplicial complexes. We  give an expression for the structural cutoff of simplicial complexes that for simplicial complexes of dimension $d=1$ reduces to the structural cutoff of simple networks. Finally  we provide a numerical analysis of the natural correlations emerging in the configuration model of simplicial complexes without structural cutoff.
 \end{abstract}


\maketitle
\section{Introduction}

{Network  theory  has  been  successful  over  the  last  fifteen  years  in  characterizing  social, }
technological   and  biological  networks.  Nevertheless,  the  increasingly  large  data  sets 
available  in  the  field   require  the   development  of  more  sophisticated  models  of networks \cite{Interdisciplinary}
such  as  multilayer  networks \cite{PhysReports,Kivela} and  generalized  network  structures \cite{Newman1,Newman2}.  In  particular  a  wide 
variety  of  networks,  including  brain networks \cite{Bassett,Hess}, social and collaboration  networks \cite{Newman_social},  immune   networks \cite{Annibale}, tagged social networks \cite{Newman1,Newman2} and ``folksonomies"  \cite{Barrat,Cattuto}, can be modeled by simplicial complexes  \cite{ENG,PRE,CQNM,Flavor,Dima}.
Therefore progress in understanding and modelling simplicial complexes has a variety of applications, ranging from brain research and  data mining \cite{Bassett, Vaccarino1,Vaccarino2},  to recommendation algorithms \cite{Zhang}, characterization of dynamical processes \cite{Mason}, and  inference of  missing links \cite{Carlo1}. 

Simplicial  complexes  are  a  generalization   of  networks  constructed  using  not only nodes 
and  links  (that  are  respectively  simplices  of  dimension  zero  and  one)  but  also   using  
triangles  (simplices   of  dimension $d=2$),  tetrahedra  (simplices  of  dimension $d=3$)  and 
higher  dimensional  simplices. Using a theoretical physics terminology, simplicial complexes describe the {\em many-body} interactions between two or more nodes.

Simplicial complexes are emerging as a new tool to describe complex networks with large clustering coefficient and  abundant number of short loops that  are not easily treatable by traditional  statistical mechanics approaches. The presence of many short loops in real network datasets has often been recognized as a signature of a hidden geometry of networks \cite{MarianM1,MarianM2}. Simplicial complexes  are ideal mathematical objects for discretizing geometry as is demonstrated by their wide use in the context of quantum gravity \cite{Oriti,Oriti_PRL,CDT1,CDT2} and therefore they can also open new scenarios in uncovering the hidden geometry of complex networks.  

Finally   simplicial complexes  constitute the network-like structure that allows for the topological analysis of network datasets. The area of network topology is currently the subject of increasing interest, with recent investigations characterizing brain networks and network dynamics \cite{Bassett,Hess,Vaccarino1,Vaccarino2,Mason} providing results so far unobtainable through other network approaches. 

For all these reasons it has become necessary to build null models for simplicial complexes using equilibrium and non-equilibrium approaches. Interestingly, extending our knowledge of static  and growing network models \cite{Hamiltonian,Kartik2009,Charo1,Charo2,EPL2008,PREGB_2009,Bianconi2008,Kartik2010,Garlaschelli1,Diaz1,Diaz2,Dorogovtsev} to simplicial complexes might reveal the role of the dimensionality of simplicial complexes in determining their structure.
 
Recently a new framework for non-equilibrium growing simplicial complexes has been formulated \cite{ENG,PRE,CQNM,Flavor}. This framework is able to generate in one limit complex manifolds of dimension $d$, in another limit complex networks growing with preferential attachment. Interestingly it has been observed that for dimension $d>2$ growing manifolds    are  scale-free, because the increase of the dimensionality of simplicial complexes over $d=2$ allows for the emergence of an efficient preferential attachment \cite{CQNM}. Interestingly, in this context it has also been shown that  simplicial complexes growing by uniform attachment of simplicies generate scale-free networks for $d\geq 2$ \cite{Flavor}.

The formulation of  equilibrium models of simplicial complexes is currently a hot topic in graph theory and pure mathematics \cite{Kahle,Farber1,Farber2}. Recently, exponential random simplicial complexes have also been attracting the attention of physicists and network scientists \cite{Dima}.

Here we develop an equilibrium statistical mechanics approach for simplicial complexes of dimension $d$. In particular we consider simplicial complexes formed exclusively by $d$ dimensional simplices. We characterizes their structure with the  generalized degree introduced in \cite{CQNM,PRE,Flavor} and defined as the number of $d$-dimensional simplices incident to a given $\delta$-dimensional face. Moreover we  treat in detail the configuration model and the canonical ensemble of simplicial complexes respectively with given generalized degree of the nodes and with expected generalized degrees of the nodes. The configuration model for simplicial complexes generalizes the configuration model for simple networks \cite{Kartik2009,Charo1,Charo2} and the hypergraph model proposed in \cite{Newman1}. The canonical ensemble is instead to be related to exponential random simplicial complexes \cite{Dima}.

These ensembles can be treated using statistical mechanics arguments that are able to characterize their relation.
Already in the context of  simple networks one can distinguish between  micro-canonical and canonical conjugated network ensembles, which enforce respectively hard or soft constraints \cite{Kartik2009,EPL2008,PREGB_2009,Bianconi2008,Kartik2010}. For example the configuration model enforcing a given degree sequence and the exponential ensemble enforcing the expected degree sequence  are respectively the micro-canonical and the canonical conjugated network ensembles.
Similarly, here we show that  the configuration model of simplicial complexes is the micro-canonical ensemble conjugated to  the canonical ensemble given by  the exponential random simplicial complex. Interestingly, here we show that  the two ensembles treated in this paper enforce an extensive number of constraints and therefore, as already noted in the context of simple networks \cite{Kartik2009}, they are not asymptotically equivalent.

The entropy of these ensembles, that has a number of applications in network analysis, and network inference \cite{Theta,Theta_jacopo} is here calculated analytically. From the entropy of the configuration model of simplicial complexes the asymptotic combinatorial  formula for the number of simplicial complexes in the ensemble  is derived. This formula generalizes the Canfield-Bender formula for the number of networks in the sparse configuration model \cite{CBender}.
When characterizing the properties of these ensembles, a special role is played by their structural cutoff that is  the maximum generalized degree that guarantees the absence of correlations between the generalized degrees of the nodes in the simplicial complex.
In any simplicial complex of dimension $d>1$, the  structural cutoff is larger than the structural cutoff of simple networks \cite{Marian_cutoff}. 
In absence of the structural cutoff simplicial complexes show relevant degree correlations analyzed here by numerical simulations. These results extend the known results observed in the canonical ensemble of simple networks \cite{Garlaschelli1}.

The paper is structured as follows: in Sec II we introduce simplicial complexes and the generalized degree of their nodes; in Sec III we treat the canonical ensemble of simplicial complexes enforcing a given sequence of expected generalized degrees of the nodes; in Sec IV we treat by statistical mechanics methods the configuration model of simplicial complexes with given sequence of generalized degrees of the nodes; in section V we discuss the natural correlations observed in our numerical realizations of the  the configuration model  of simplicial complexes; finally in Sec VI we give the conclusions.

\section{Simplicial complexes and generalized degrees }
\subsection{Simplicial complexes of general dimension $d$}

 A $d$-dimensional simplex is formed  by  a set of $(d+1)$ interacting nodes, and includes all  the  subsets of $\delta+1$  nodes (with $\delta<d$) which are called  the $\delta$-dimensional faces of the simplex. 
A simplicial complex of dimension $d$ is formed by simplices of dimension at most equal to  $d$ glued along their faces.

As mathematical objects simplicial complexes  are distinct from  hypergraphs \cite{Newman1,Newman2}, the difference being that simplicial complexes include all the subsets of a given simplex. Nevertheless in most of the interesting network science applications the terms simplicial complex and hypergraph might be used to indicate the same type of network data.
 
 Here we consider $d$-dimensional simplicial complexes of $N$ nodes formed exclusively by $d$-dimensional simplices.
We indicate with ${\cal Q}_d(N)$ the  set of all possible and distinct $d$-dimensional simplices in a $d$-dimensional simplicial complex of $N$ nodes   while we indicate with ${\cal S}_{d,\delta}$ the set of all $\delta$-dimensional simplices present in a given  $d$-dimensional simplicial complex.
The simplicial complexes that we consider in this paper are fully identified once the adjacency tensor $\bf a$ is fully specified.
The adjacency tensor ${\bf a}$ has elements $a_{\alpha}=0,1$ indicating for each possible $d$-dimensional simplex $\alpha\in {\cal Q}_{d}(N)$ if the simplex is present ($a_{\alpha}=1$) or absent ($a_{\alpha}=0$) in the simplicial complex, i.e.
\bea
a_{\alpha}=\left\{\begin{array}{ccc} 1 & {\mbox{if}}& \alpha\in {\cal S}_{d,d}\nonumber \\
0&&\mbox{otherwise} \end{array}\right..
\eea 
The {\em generalized degrees} \cite{CQNM,Flavor} are relevant structural properties of  simplicial complexes.
The generalized degree  $k_{d,\delta}(\alpha)$ of a  $\delta$-dimensional face (or $\delta$-face)  $\alpha$ of the $d$-dimensional simplicial complex  quantifies the number of $d$-dimensional simplices incident to the $\delta$-face $\alpha$. 
The generalized degree $k_{d,\delta}(\alpha)$ can be defined in terms of the adjacency tensor ${\bf a}$ as
\bea
k_{d,\delta}(\alpha)=\sum_{\alpha'\in {\cal Q}_d (N)|\alpha'\supseteq \alpha }a_{\alpha'}.
\eea
The generalized degrees are not independent on each other. In fact  the generalized degree of a $\delta-$face $\alpha$ is   related to the generalized degree of the $\delta'$-dimensional faces incident to it, with $\delta'>\delta$, by the simple combinatorial relation 
\bea
k_{d,\delta}(\alpha)=\frac{1}{\left(\begin{array}{c}d-\delta\\ \delta'-\delta\end{array}\right)}\sum_{\alpha'\in {\cal Q}_d (N)|\alpha'\supseteq \alpha}k_{d,\delta'}(\alpha')
\eea
Moreover, since every $d$-dimensional simplex belongs to $\left(\begin{array}{c}d+1\\ \delta+1\end{array}\right)$ $\delta$-dimensional faces, in a simplicial complex with $M$ $d$-dimensional simplices we have
\bea
\sum_{\alpha\in {\cal S}_{d,\delta}}k_{d,\delta}(\alpha)=\left(\begin{array}{c}d+1\\ \delta+1\end{array}\right)M.
\eea
In this paper we focus specifically on the generalized degree of the nodes $r=1,2,\ldots N$ given by 
\bea
k_{d,0}(r)=\sum_{\alpha'\in {\cal Q}_d (N)|\alpha'\supset r  }a_{\alpha'}.
\label{kd0r}
\eea
The generalized degree of the node indicates the number of $d-$dimensional simplices incident to each node $r$.
Clearly, since the simplicial complexes under investigation  are only formed by $d$-dimensional simplices, the generalized degree of the nodes satisfy
\bea
\sum_{r=1}^Nk_{d,0}(r)=(d+1) M,
\label{Md}
\eea
where $M$ are the number of $d$-dimensional simplices in the simplicial complex.
The generalized degree $k_{d,0}(r)$ of the nodes will play a crucial role in this paper because we will discuss the properties of the configuration model with given generalized degree sequence of the nodes.
In the subsequent sections we will focus on the configuration model and the canonical ensemble for simplicial complexes enforcing respectively a given sequence of the  generalized degree of the nodes and given sequence of the expected generalized degree of the nodes.
We will always consider networks in which the number of simplices $M$ is of the same order of magnitude of the number of nodes 
\bea
M\propto N, 
\eea 
which is the relevant regime for most of the applications to complex networks.

\begin{figure}
\begin{center}
{\includegraphics[width=0.99\columnwidth]{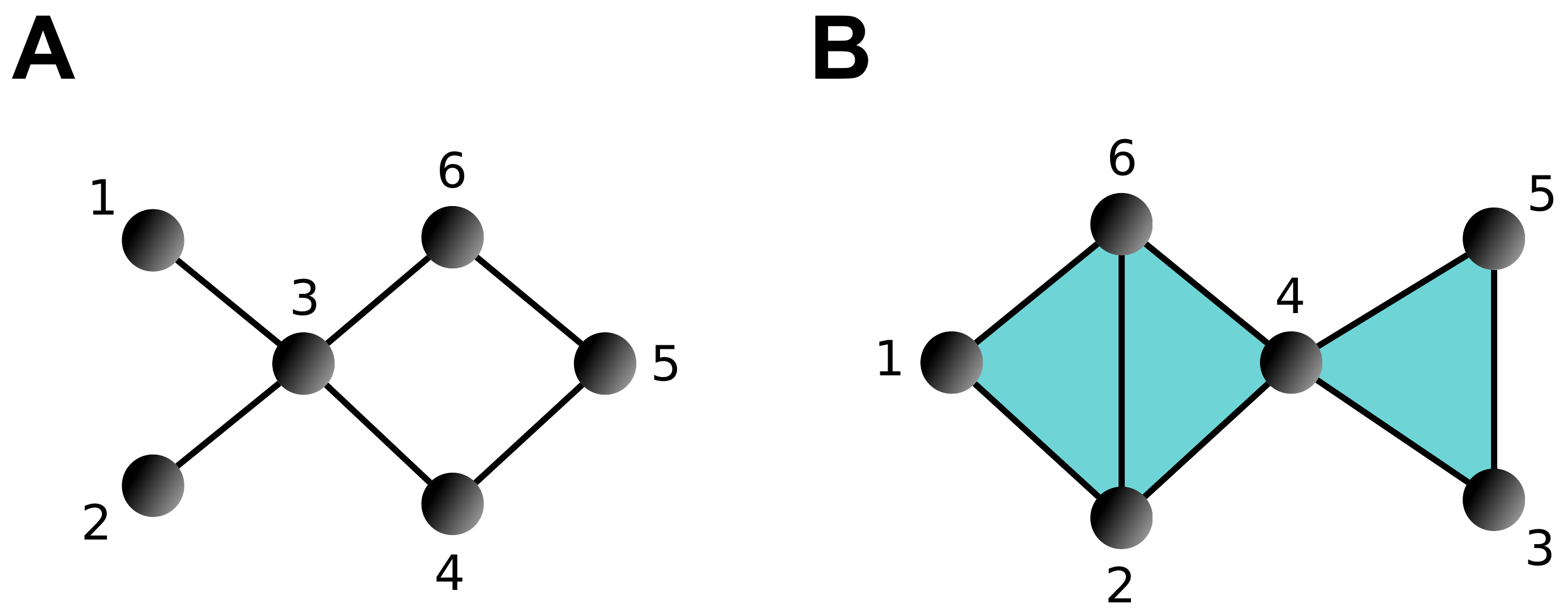}}
\end{center}
\caption{(Color online) Examples of simplicial complexes of dimension $d=1$ (panel A) and $d=2$ (panel B) are shown. Simplicial complexes of dimension $d=1$ are simple networks. Simplicial complexes of dimension $d\geq 2$  characterize interactions occurring between more than two nodes, (specifically interactions occurring between $d+1$ nodes).}
\label{fig_SC}
\end{figure}

Before discussing the properties of these ensembles,  in the following paragraphs we will characterize simplicial complexes of dimension $d=1,2$ using the generalized degrees of their $\delta$-faces (for examples of simplicial complexes in dimension $d=1,2$ see Figure $\ref{fig_SC}$).The extension to higher simplicial complexes is straightforward.
 
\subsection{Case of a simplicial complex of dimension $d=1$}
Simplicial complexes of dimension $d=1$ are formed exclusively by nodes and links (which are the $1$-dimensional simplices). The adjacency tensor of the $1$-dimensional  simplicial complex is nothing else than the adjacency matrix $\{a_{rm}\}$, with elements $a_{rm}$ indicating if the link $(r,m)$ is present or not in the network.
In this case the  generalized degree $k_{1,0}(r)$ of the nodes, simply indicate the number of links incident to the node, i.e. its degree.
 In fact we have 
 \bea
 k_{1,0}(r)=\sum_{m=1}^N a_{rm}.
 \label{k10r}
 \eea

\subsection{Case of a simplicial complex of dimension $d=2$}
Here we consider  the case of a simplicial complex of dimension $d=2$ characterizing interactions occurring between $3$ nodes, i.e. a  simplicial complex of formed exclusively  by triangles. We assume that the number of nodes in the simplicial complex is $N$.
This simplicial complex  is determined by the adjacency tensor
$\{a_{rmn}\}$ of elements $a_{rmn}=1$ if the nodes $(r,m,n)$ are linked by a triangle, and  $a_{rmn}=0$ if the nodes $(r,m,n)$ are not connected by a triangle.
The generalized degree $k_{2,0}(r)$ of node $r$ is given by 
\bea
k_{2,0}(r)=\sum_{m<n}a_{rmn},
\label{k20r}
\eea
while the generalized degree $k_{2,1}(r,m)$ of a link $(r,m)$ is given by 
\bea
k_{2,1}(r,m)=\sum_{n}a_{rmn}.
\eea
The generalized degree $k_{2,0}(r)$ of node $r$ indicates the number of triangles  incident to it, while the generalized degree $k_{2,1}(r, m)$ of the link $(r,m)$ indicates the number of triangles incident to the link.
The generalized degree of the nodes is related to the generalized degree of the links.
In fact it is easy to see that 
\bea
\hspace*{-5mm}k_{2,0}(r)=\sum_{m<n}a_{rmn}=\frac{1}{2}\sum_{m,n}a_{rmn}=\frac{1}{2}\sum_{m}k_{2,1}(r, m).
\eea
Since each triangle is incident to three nodes, we have
\bea
\sum_{r=1}^Nk_{d,0}(r)=3 M,
\label{M}
\eea
where $M$ are the number of $d$-dimensional simplices in the simplicial complex.

\section{Canonical ensemble of simplicial complexes}

\subsection{Canonical ensemble with given sequence of expected generalized degree of the nodes}

In this section we  discuss  the canonical ensemble of  simplicial complexes (also called the exponential random simplicial complex ) with given sequence of expected generalized degree of the nodes. In this ensemble of simplicial complexes  each simplicial complex $G$ is assigned a probability $P(G)$.
The entropy $S$ of the ensemble evaluates the typical number of simplicial complexes belonging to the ensemble and is  given by 
\bea
S=-\sum_{G}P(G)\ln P(G),
\label{Sentropy}
\eea
where the sum is extended to all simplicial complexes $G$ under consideration, or equivalently, over all adjacency tensors ${\bf a}$. 

The canonical ensemble is  the least biased ensemble of simplicial complexes that satisfies the constraints
\bea
\overline{k_r}=\overline{k_{d,0}(r)}=\sum_{G}P(G)\sum_{\alpha\in {\cal Q}_d (N)|r\subset \alpha}a_{\alpha}.
\label{constraints}
\eea

The canonical  ensemble is  the maximum entropy ensemble satisfying the constraints in Eq. $(\ref{constraints})$.
Therefore, in order to derive the probability $P(G)$ of a simplicial complex $G$ in the canonical  ensemble, we maximize the functional ${\cal F}$ given by 
\bea
{\cal F}&=&S+\sum_{r=1}^N\lambda_r\left[\overline{k_{r}}-\sum_{G}P(G)k_{d,0}(r)\right]\nonumber \\
&&+\mu\left[1-\sum_{G}P(G)\right],
\eea
where we have introduced, by means of the $N$ Lagrangian multipliers $\lambda_r$, the constraints  in Eq. $(\ref{constraints})$,
and by means of the Lagrangian multiplier $\mu$ the normalization constraint for  the probability $P(G)$.
Maximizing ${\cal F}$ with respect to $P(G)$, we obtain that the canonical  ensemble of simplicial complexes enforcing a given sequence of expected generalized degrees of the nodes $\{\overline{k_r} \}$,  has probability given by 
\bea
P(G)=\frac{1}{Z}e^{-\sum_r \lambda_r k_{d,0}(r)}
\label{PG1}
\eea
where  $k_{d,0}(r)$ is given by Eq. $(\ref{kd0r})$, the normalization constant $Z$ is given by 
\bea
Z=\prod_{\alpha\in{\cal Q}_{d}(N)}\Big[1 + e^{-\sum_{r \subset \alpha}\lambda_{r}}\Big].
\eea
The Lagrangian multipliers  $\{\lambda_r \}$ occurring in Eq. ($\ref{PG1}$) are fixed by the Eqs. $(\ref{constraints})$ and ($\ref{PG1}$). Substituting the expression for $P(G)$ in ($\ref{PG1}$) into $(\ref{constraints})$ we get 
\bea
\overline{k_{r}}=\sum_{\alpha \in {\cal Q}_d (N)|r\subset \alpha}\frac{e^{-\sum_{m \subset \alpha}\lambda_{m} }}{1 + e^{ -\sum_{m\subset \alpha}\lambda_{m}}}.
\label{kr} 
\eea
The probability $p_{\alpha}$ that each $d$-dimensional simplex $\alpha\in {\cal Q}_{d}(N)$ is in the simplicial complex, is given by 
\bea
p_{\alpha}=\sum_G P(G)a_{\alpha}=\frac{e^{-\sum_{r \subset \alpha}\lambda_{r} }}{1 + e^{ -\sum_{r\subset \alpha}\lambda_{r}}} 
\label{palpha0}
\eea
Interestingly for this ensemble,  the probability $P(G)$ can be written as a product of the marginal probabilities for the individual $d$-dimensional simplices $p_{\alpha}$, i.e.
\bea
P(G)=  \prod_{\alpha\in{\cal Q}_{d}(N)}\Big[p_{\alpha}^{a_{\alpha}} (1- p_{\alpha})^{1-a_{\alpha}}\Big]
\label{PG}
\eea
Consequently the entropy $S$ can be written as 
\bea
S= -\sum_{\alpha\in{\cal Q}_{d}(N)}\Big[p_{\alpha}\ln{p_{\alpha}} +(1-p_{\alpha})\ln(1-p_{\alpha})\Big]
\label{S2}
\eea

\subsection{The canonical ensemble of simplicial complexes with  structural cutoff}
As long as the maximum generalized degree of the nodes is smaller than the structural cutoff, the probabilities $p_{\alpha}$ can be expressed as the normalized product of the generalized degrees of the nodes belonging to $\alpha$.
In fact assuming $e^{-\lambda_r}\ll1$, the probability $p_{\alpha}$ given by Eq. $(\ref{palpha0})$ can be approximated by 
\bea
p_{\alpha}\simeq\prod_{r\subset \alpha, \alpha  \in {\cal Q}_d (N)}e^{-\lambda_r}.
\eea
Eq. $(\ref{kr})$ can now be simplified and rearranged to give an explicit expression for $e^{-\lambda_r}$ in terms of $k_{r}$ and the other Lagrangian multipliers:
\bea
e^{-\lambda_r}=\overline{k_{r}}\frac{d!}{(\sum_m e^{-\lambda_m})^d}.
\label{uno}
\eea
Note that in this last expression we made the following approximation 
\bea
\sum_{m_1<m_2<\ldots<m_{d+1}}\prod_{j=1}^{d+1}e^{-\lambda_{m_{j}}}\simeq \frac{1}{d!} \left(\sum_m e^{-\lambda_m}\right)^d,
\eea
valid in the limit in which the number of nodes $N$ is large, i.e. $N\gg1$ and $e^{-\lambda_r}\ll1$.
Summing  over all the nodes of the simplicial complex, we get
\bea
\sum_r e^{-\lambda_r}=\Big(\avg{\overline{k}}N d!\Big)^{1/(d+1)}.
\label{due}
\eea
Finally, combining Eq.$(\ref{uno})$ and Eq.$(\ref{due})$ we get
\bea
e^{-\lambda_r}=\overline{k_{r}}\left[\frac{d!}{(\avg{\overline{k}}N)^d}\right]^{1/(d+1)}.
\label{tre}
\eea
Using this result we get the simplified expression for the probability $p_{\alpha}$ of the $d$-dimensional simplex $\alpha$, given by 
\bea
p_{\alpha}=d!\frac{\prod_{r\subset \alpha}\overline{k_{r}}}{(\avg{\overline{k}}N)^d},
\label{pa}
\eea
where $\alpha \in {\cal Q}_d (N)$.
This expression is valid as long as $e^{-\lambda_r}$ (given by Eq. $(\ref{tre})$) satisfies the hypothesis $e^{-\lambda_r}\ll1$. 
This implies that the maximum generalized degree of the nodes  $K_{max}$ should be much smaller than the structural cutoff $K_d$ for simplicical complexes, i.e.
\bea
K_{max}\ll K_d=\left[\frac{\left({\avg{\overline{k}}N}\right)^{d}}{d!}\right]^{1/(d+1)}
\label{Kd2}
\eea
Interestingly the cutoff $K_d$ for the present ensemble of simplicial complexes scales like $N^{d/(d+1)}$, i.e. it is increasing with an exponent that is larger for larger dimensions $d$.

This regime is the regime in which there are no correlations between the generalized degrees of the nodes. Moreover in this regime only few links can be incident to more than one $d$-dimensional simplex.
In fact, given the expression for $p_{\alpha}$ provided by Eq. $(\ref{pa})$, it is possible to evaluate in this ensemble the expected generalized degree of the link $\overline{k_{d,2}(r,m)}$ for $d>2$.
This is given by 
\bea
\overline{k_{d,1}(r,m)}=\sum_{\alpha|(r,m)\subset \alpha}p_{\alpha}=d\frac{\overline{k_{r}}\ \overline{k_{m}}}{\avg{\overline{k}}N}.
\eea 
Therefore only the pairs of nodes $(r,m)$ with generalized degree of the nodes $\overline{k_{r}},\overline{k_{m}}\gg N^{1/2}$ and $\overline{k_{r}},\overline{k_{m}}\ll N^{d/(d+1)}$ are likely to be incident to more than one $d$-dimensional simplex.

\subsection{The canonical ensemble of simplicial complexes  of dimension $d=1$}

For $d=1$ our construction  of the  canonical ensemble of simplicial complexes  for networks reduces to the canonical ensemble (exponential ensemble) of networks \cite{Hamiltonian} with  given expected degree sequence. 
The probability $P(G)$ of a given $1$-dimensional simplicial complex (i.e. network) specified by the adjacency tensor $\{a_{rm}\}$ is given by Eq. $(\ref{PG1})$  given in this case by 
\bea
P(G)= \frac{1}{Z} e^{-\sum_{r} \lambda_{r} k_{1,0}(r)}
\eea
where $k_{1,0}(r)$ is given by Eq. $(\ref{k10r})$ and the normalization constant $Z$ is given by 
\bea
Z=\prod_{r<m}\left(1+e^{-\lambda_r-\lambda_m}\right).
\eea
The Lagrangian multipliers $\lambda_r$ are fixed by the condition
\bea
\overline{k_r}=\overline{k_{1,0}(r)}=\sum_{m} p_{rm}
\eea
with $p_{rm}$ indicating the probability that the link between the nodes $r,m$ is present in the network. The probability $p_{rmn}$ are  given by 
\bea
p_{rm}=\frac{e^{-(\lambda_{r} + \lambda_{m} })}{1 + e^{ -(\lambda_{r} + \lambda_{m} )}}. 
\eea
The probability $P(G)$ of a simplicial complex $G$ in this canonical ensemble can be expressed as a  product of the marginal probabilities for the individual links:
\bea
P(G)=  \prod_{r<m}\Big[p_{rm}^{a_{rm}} (1- p_{rm})^{1-a_{rm}}\Big].  
\eea
Therefore the entropy $S$ of the ensemble is given by 
\bea
S=-\sum_{r<m}\left[p_{rm}\ln p_{rm}+(1-p_{rm})\ln(1-p_{rm})\right].
\eea
Finally in presence of the structural cutoff on the generalized degree of the nodes, i.e. if the maximal generalized degree of the nodes $K_{max}$ satisfies 
\bea
K_{max}\ll K_1=\left(\avg{\overline{k}}N\right)^{1/2},
\label{Kd1p}
\eea
the probabilities $p_{rm}$ take a simple factorized expression given by 
\bea
p_{rm}=\frac{\overline{k_r}\ \overline{k_m}}{\avg{\overline{k}}N}.
\eea
We note here that the structural cutoff of simplicial complexes of dimension $d=1$ given by Eq. $(\ref{Kd1p})$ reduces to the structural cutoff of simple networks \cite{Marian_cutoff} as expected.

\subsection{The canonical ensemble of simplicial complexes of dimension $d=2$}

In this subsection we summarize the results for the case of a canonical ensemble of two dimensional simplicial complexes where we constrain the expected generalized degree of the nodes to be $k_{2,0}(r)$.
The probability $P(G)$ of a given simplicial complex specified by the adjacency tensor $\{a_{rmn}\}$ which is given by Eq. $(\ref{PG1})$ that reads for this case
\bea
P(G)= \frac{1}{Z} e^{-\sum_{r} \lambda_{r} k_{2,0}(r)}
\eea
where $k_{2,0}(r)$ is given by Eq. $(\ref{k20r})$ and the normalization constant $Z$ is given by 
\bea
Z=\prod_{r<m<n}\left(1+e^{-\lambda_r-\lambda_m-\lambda_n}\right).
\eea
The Lagrangian multipliers $\lambda_r$ are fixed by the condition
\bea
\overline{k_r}=\overline{k_{2,0}(r)}=\sum_{m<n} p_{rmn}
\eea
with $p_{rmn}$ indicating the probability that the triangle between the nodes $r,m,n$ is present in the simplicial complex, which is given by 
\bea
p_{rmn}=\frac{e^{-(\lambda_{r} + \lambda_{m} + \lambda_{n}})}{1 + e^{ -(\lambda_{r} + \lambda_{m} + \lambda_{n})}}. 
\eea
The probability $P(G)$ of a simplicial complex $G$ in this canonical ensemble can be expressed as a  product of the marginal probabilities for the individual triangles:
\bea
P(G)=  \prod_{r<m<n}\Big[p_{rmn}^{a_{rmn}} (1- p_{rmn})^{1-a_{rmn}}\Big].  
\eea
Therefore the entropy $S$ of the ensemble is given by 
\bea
S=-\sum_{r<m<n}\left[p_{rmn}\ln p_{rmn}+(1-p_{rmn})\ln(1-p_{rmn})\right].
\eea
Finally in presence of the structural cutoff on the generalized degree of the nodes, i.e. if the maximal generalized degree of the nodes $K_{max}$ satisfies
\bea
K_{max}\ll K_2=\left(\frac{\avg{\overline{k}}N}{\sqrt{2}}\right)^{2/3}
\eea
the probabilities $p_{rmn}$ take a simple factorized expression given by 
\bea
p_{rmn}=2\frac{\overline{k_r}\ \overline{k_m}\ \overline{k_n}}{(\avg{\overline{k}}N)^2}.
\label{prmn2}
\eea
Here the structural cutoff $K_2$ scales like $N^{2/3}$. It is therefore much larger than the structural cutoff for simple networks.

We note  that this model is to be related with the model of tagged social networks represented by hypergraphs  presented in Ref. \cite{Newman1,Newman2}. Nevertheless it differs with respect to the cited work because in the present work the  three nodes linked in a given $2$-dimensional simplex represent   the same type of nodes. This difference is responsible for the factor two present in the right hand side of Eq. $(\ref{prmn2})$.

\subsection{Generation of simplicial complexes by the canonical ensemble}

For generating the canonical ensemble of $d$-dimensional simplicial complexes with expected generalized degree sequence of the nodes $\{\overline{k_r}\}$ with $r=1,2,\ldots N$, we propose the following algorithm:
\begin{itemize}
\item[(a)] Calculate the probabilities $p_{\alpha}$ of any $d$-dimensional simplex $\alpha\in {\cal Q}_d(N)$ given by Eq. $(\ref{palpha0})$ in absence of the structural cutoff $K_d$ or by Eq. $(\ref{pa})$ in presence of the structural cutoff $K_d$.
\item[(b)]
Draw every possible $d$-dimensional simplex $\alpha\in {\cal Q}_d(N)$ with probability $p_{\alpha}$.
\end{itemize}

\section{The  configuration model of simplicial complexes}

\subsection{The configuration model of simplicial complexes with given generalized degree of the nodes}
The configuration model of simplicial complexes with given sequence of the generalized degrees of the nodes $\{k_r\}$ is the ensemble 
in which we assign the  same probability to each simplicial complex with the generalized degrees of the nodes satisfying $k_{d,0}(r)=k_r$ for every  node $r$.
The construction of simplicial complexes is allowed only if the generalized degree sequence of the nodes is graphical, i.e. if at least one simplicial complex can be constructed with it.  For simple networks, i.e. for simplicial complexes of dimension $d=1$, the conditions that a degree sequence must satisfy in order to be graphical have been fully identified \cite{M,Charo1}. For simplicial complexes we know that at least the generalized degree of the nodes must satisfy Eq. $(\ref{Md})$. In practice, it will often be useful to start from sequences of generalized degree of the nodes occurring in real  datasets which are by definition graphical. This will be recommended in order to construct a randomized simplicial complex that will provide a null model to the real dataset.

\begin{figure*}
\begin{center}
{\includegraphics[width=1.99\columnwidth]{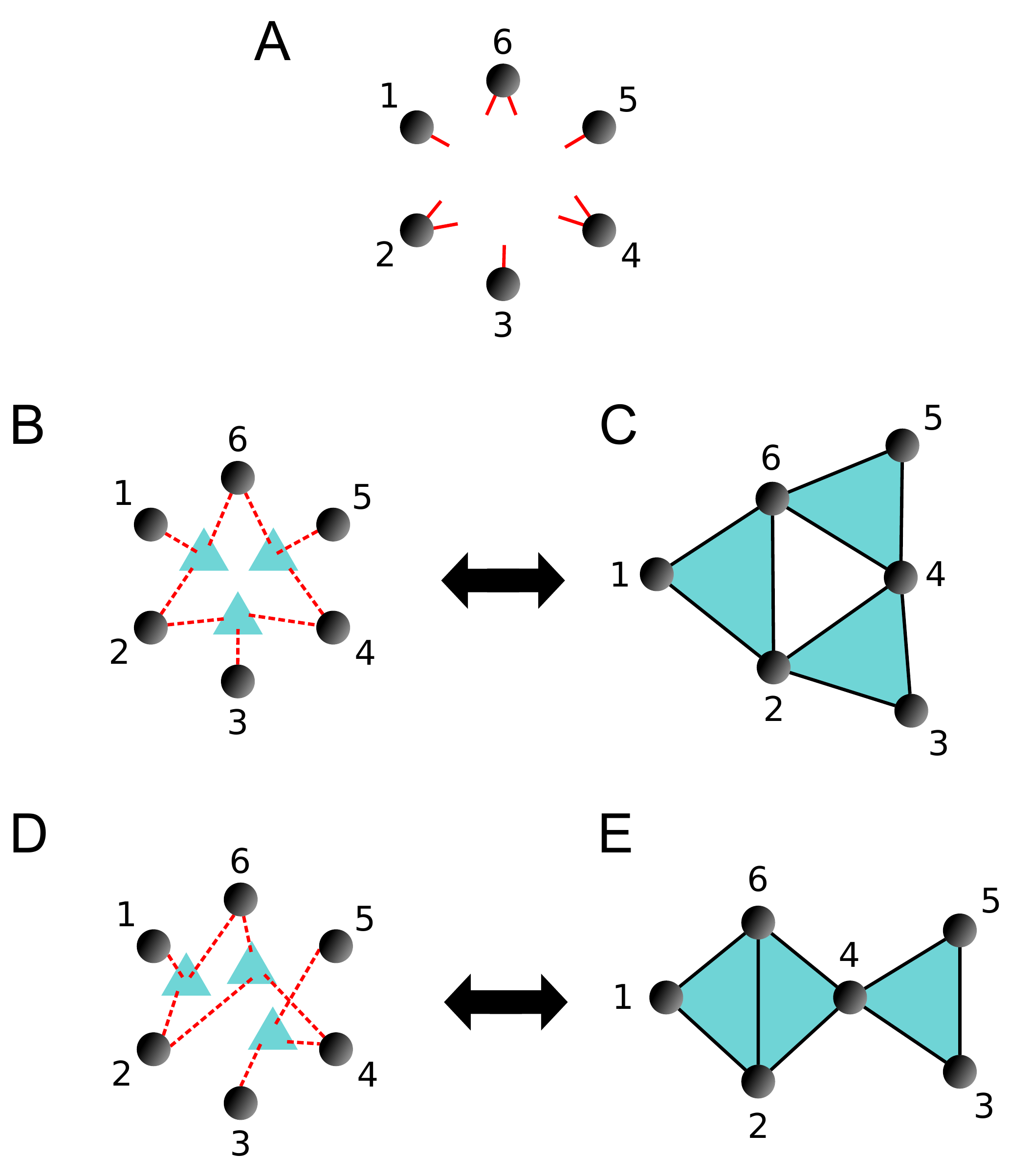}}
\end{center}
\caption{(Color online) The figure shows the construction of two different $d=2$ dimensional simplicial complexes belonging to the same configuration model of simplicial complexes. In panel A the $N=6$ nodes are shown together with stubs indicating their generalized degree. In panel B triples of stubs are matched together to form $2$-dimensional simplices. In figure C the corresponding simplicial complex is visualized.
In panels D-E a different matching of the stubs is shown together with its corresponding simplicial complex. As is evident from the figure, a given generalized degree sequence of the nodes can give rise to different simplicial complexes. The logarithm of the total number ${\cal N}$ of simplicial complexes that can be constructed from a given generalized degree sequence of the nodes, is the Gibbs entropy $\Sigma$ of the configuration model. }
\label{fig_conf}
\end{figure*}

The configuration model enforcing a given graphical sequence of the generalized degrees of the nodes, assigns  to each $d$-dimensional simplicial complex $G$ formed exclusively by $d$-dimensional simplexes  the probability 
\bea
P(G)=\frac{1}{{\cal N}}\prod_{r=1,N}\delta(k_r, k_{d,0}(r)).
\label{Pgmicro}
\eea
Here ${\cal N}$ is the number of simplicial complexes with the given graphical sequence of generalized degree of the nodes $\{k_r\}$  given by 
\bea
{\cal N}=\sum_{G}\prod_{r=1,N}\delta(k_r,k_{d,0}(r)).
\eea
In Figure $\ref{fig_conf}$ we show how from a given graphical sequence of generalized degree of the nodes it is possible in general to construct different simplicial complexes. 

\subsection{Generation of the simplicial complexes by the configuration model  }
\label{conf_al}

In this paragraph  we generalize the algorithm for the configuration model of networks with given degree sequence to the configuration model of $d$-dimensional simplicial complexes with given sequence $\{k_r\}_{r\leq N}$ of the generalized degrees of the nodes. For describing this algorithm, we will use a set of $M$ auxiliary {\em factor nodes} $\mu=1,2\ldots, M$, with $M$ satisfying 
\bea
\sum_{r=1}^Nk_r=(d+1)M.
\eea

\begin{figure}
\begin{center}
{\includegraphics[width=0.99\columnwidth]{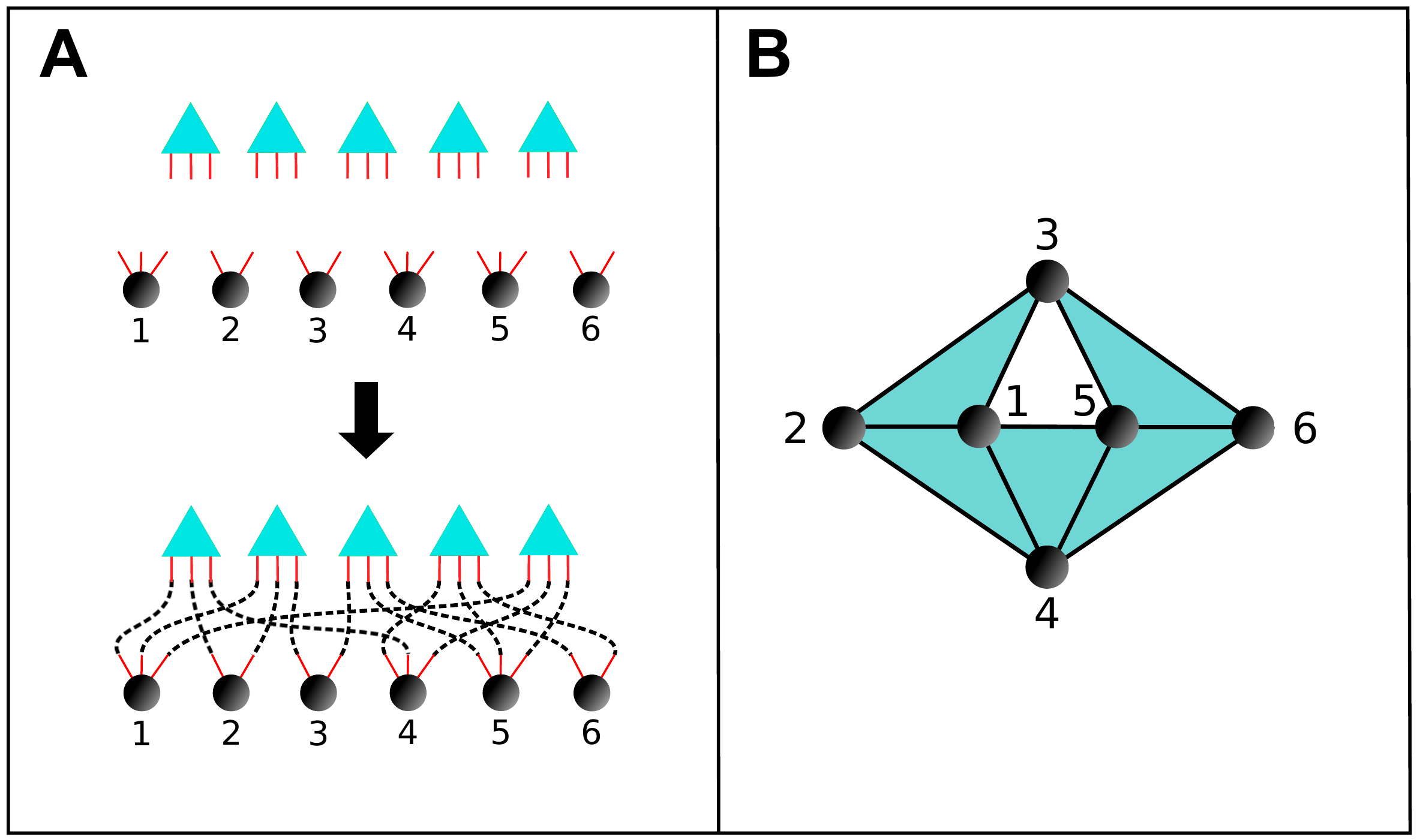}}
\end{center}
\caption{(Color online) A scheme representing the algorithm for the construction of the configuration model is shown for the case $d=2$. Panel A represents the Steps (i)-(ii)-(iii). To each node $r$  with $r=1,2\ldots, N=6$  we assign $k_r$ stubs. The nodes are represented with black circles. A set of $M$ auxiliary factor nodes (cyan triangles) is considered. Each factor node has  $d+1$ stubs. Subsequently an allowed matching of the stubs is found. Panel B shows how from the matching of the stubs we can construct a simplicial complex by adding a simplex between all of the nodes connected to a common factor node in panel A.}
\label{fig2}
\end{figure}

The algorithm is described in Figure $\ref{fig2}$ in the case $d=2$ and proceeds as follows:
\begin{itemize}
\item[(i)] Initially,  $k_r$ stubs are placed on each node $r=1,2,\ldots, N$. Additionally,   $d+1$ stubs are placed on each  auxiliary factor node $\mu=1,2,\ldots M$. Initially each stub is unmatched.
\item[(ii)] A set of $d+1$ unmatched random stubs of the nodes is chosen with uniform probability. Without losing generality we assume that the  stubs belong to the set of nodes $(r_1,r_2,\ldots, r_{d+1})$.
\item[(iii)] If the  nodes $(r_1,r_2,\ldots, r_{d+1})$  are all distinct and  no factor node $\mu$ is already matched with the set of nodes $(r_1,r_2,\ldots, r_{d+1})$, we match the $d+1$ stubs of an unmatched   random factor node  to the nodes $(r_1,r_2,\ldots, r_{d+1})$. Otherwise we start again from Step (i).
\item[(iv)] If all the stubs are matched we construct the simplicial complex by placing a simplex between the nodes connected to each auxiliary factor node.
\end{itemize}
In Figure $\ref{fig2}$ we show an example of the possible matching of the stubs of nodes and factor nodes and the consequent construction of the simplicial complex.

\begin{figure}
\begin{center}
{\includegraphics[width=0.99\columnwidth]{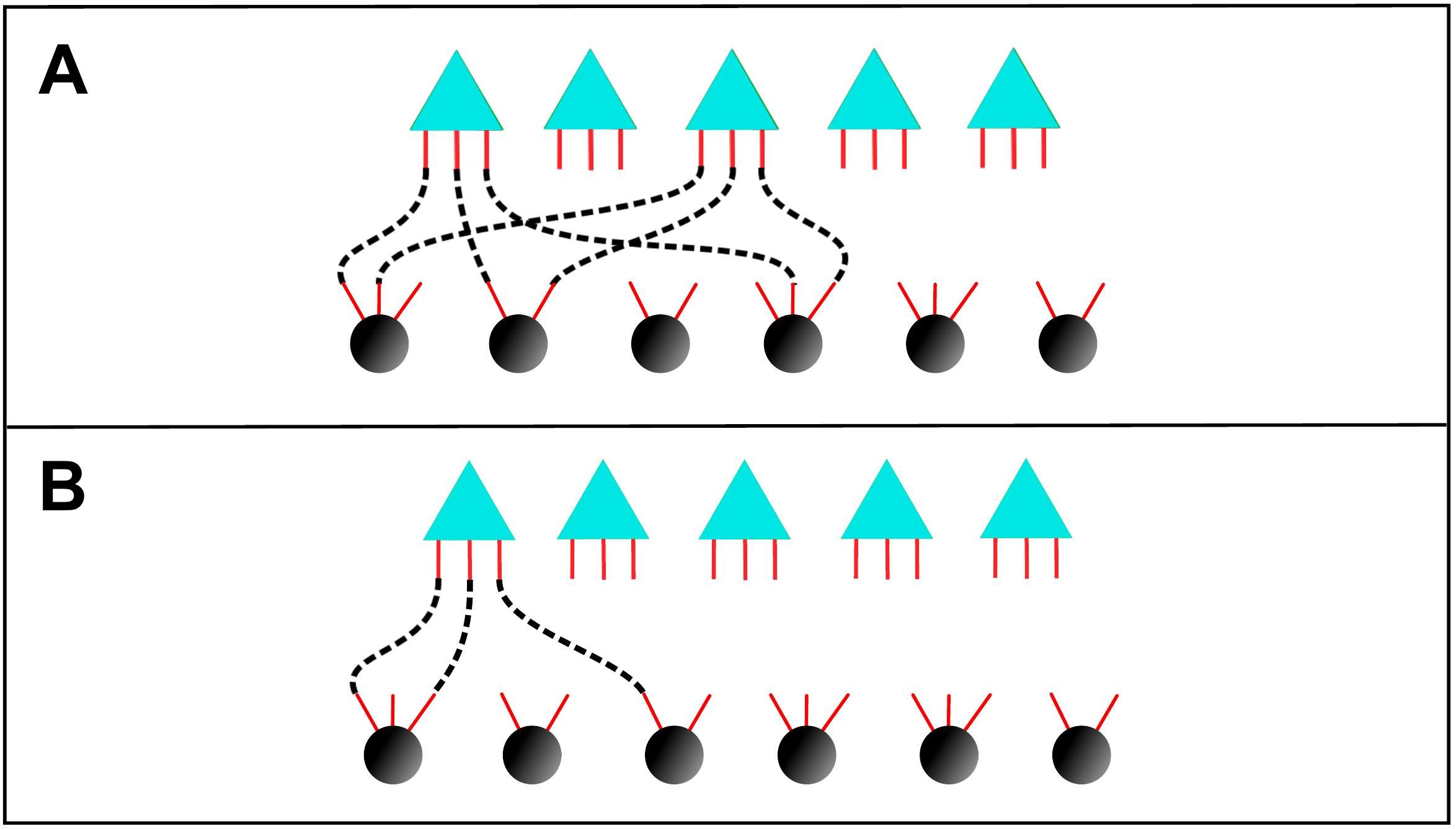}}
\end{center}
\caption{(Color online)  Two examples of forbidden moves are shown. In panel A the same set of nodes $(r_1,r_2,\dots, r_{d+1})$ is selected more than once to form a simplex. In panel B the set of nodes $(r_1,r_2,\ldots, r_{d+1})$ selected to form a simplex is not formed by $d+1$ distinct nodes. Here the forbidden moves  are shown for the configuration model of simplicial complexes of dimension $d=2$.}
\label{fig3}
\end{figure}

The step (iii) rejects moves that are forbidden. These moves are described in Figure $\ref{fig3}$. This rejection procedure guarantees that there are no spurious correlations in the structure of the simplicial complex, but for broad distribution of the generalized degrees of the nodes it might significantly slow down the algorithm.
In the context of the configuration model, more sophisticated algorithms have been proposed in Ref. \cite{Charo1,Charo2} and we believe that along these lines it could also be possible to optimize the code for the case of simplicial complexes in the future.

Here, when numerically implementing the algorithm (see Supplementary Material \cite{SM} for the codes generating random simplicial complexes in  $d=1,d=2,d=3$), we  have chosen to allow a rejection of a small number $n_F$ of forbidden moves. Therefore we have modified the above algorithm by substituting step $(iii)$ with :
\begin{itemize}
\item[(iii)-a] If the  nodes $(r_1,r_2,\ldots, r_{d+1})$  are all distinct and  no factor node $\mu$ is already matched with the set of nodes $(r_1,r_2,\ldots, r_{d+1})$, we match the $d+1$ stubs of an unmatched   random factor node  to the nodes $(r_1,r_2,\ldots, r_{d+1})$. 
\item[(iii)-b] If  the  nodes $(r_1,r_2,\ldots, r_{d+1})$  are not all distinct or  a factor node $\mu$ is already matched with the set of nodes $(r_1,r_2,\ldots, r_{d+1})$ we update a variable $n_x$ that counts how many  similar events  have occurred so far.
If $n_x \leq n_F$   we do not accept the move and we go back to Step (ii), if $n_x>n_F$ we go back to the initial  Step (i).
\end{itemize}
This algorithm reduces to the one described before when $n_F=1$, and when $n_F\ll N$ it speed up significantly the code, without altering significantly the properties of the simplicial complexes.

\subsection{Relation with bipartite network models}
Bipartite networks are formed by a set of nodes $r=1,2,\dots, N$ and a set of factor nodes (groups) $\mu=1,2,\ldots, P$ where links only join a node with a factor node (a group).
A given bipartite network has adjacency matrix ${\bf A}$ with elements $A_{r,\mu}=1$ if the node $r$ belongs to group  $\mu$, and $A_{r,\mu}=0$ otherwise.
The bipartite network might for example describe   a network formed by scientists (the nodes) and by scientific papers (the groups) where each paper is connected to the set of its authors. Similar models have been proposed for social networks \cite{Newman_social} and for immune   networks \cite{Annibale}.
Each group $\mu$  of a bipartite networks can be related to a simplicial complex constructed by joining all the  nodes connected to a common factor node (group). In particular if all the factor nodes have the degree equal to $d$, all these simplicial complexes are $d$ dimensional.
Therefore  it is  important to discuss here  the relation between the configuration model for simplicial complexes and the ensemble of bipartite networks in which we fix the degree sequence of the nodes and of the factors nodes to be 
 \bea
 \sum_{\mu}A_{r \mu}&=&k_r,\nonumber \\
 \sum_{r=1}^NA_{r \mu}&=&d+1.
 \label{bip}
 \eea
The differences between the configuration model for bipartite networks with constraints given by Eq. $(\ref{bip})$ and the configuration model for simplicial complexes are:
\begin{itemize}
\item[(1)] In the bipartite network it is possible to observe more than one factor node connecting the same set of nodes.
\item[(2)] In the bipartite network the factor nodes are labelled. 
\end{itemize}
An example illustrating these differences is that of a bipartite network between authors and papers co-authored by three authors and the corresponding simplicial complex describing the collaboration network between the authors. The difference between these two datasets is that bipartite networks distinguish between situations where three authors write only one or several papers together, and they also distinguish between papers with the same three authors (i.e. the papers are labelled).  In contrast simplicial complexes indicate only whether a given set of three authors have co-authored at least one paper together, independently on the paper title and content.

\subsection{Canonical ensemble conjugated to the configuration model of simplicial complexes}
The configuration model for simplicial complexes enforcing the generalized degree sequence of the nodes $\{k_r\}$ and the canonical ensemble of simplicial complexes enforcing the expected generalized degree of the nodes $\{\overline{k_r}\}$ are {\it conjugated network ensembles} when $\overline{k_r}=k_r$ for every node $r$.
The configuration model can also be called the micro-canonical ensemble conjugated to the  canonical ensemble of simplicial complexes.
This terminology is borrowed from statistical mechanics that treats ensembles of dynamical systems with given energy (micro-canonical  ensemble) or with given expected (average) energy (canonical ensemble). 
In statistical mechanics these two ensembles are thermodynamically equivalent, i.e. their statistical properties are the same when one considers systems formed by large number of particles as for example  a gas of molecules. 
For network ensembles, the most fundamental example of conjugated micro-canonical and canonical ensembles are the Erd\"os and Renyi random graphs in which we fix the total number of links (micro-canonical ensemble) and the random graph in which we fix  the average number of links (canonical ensemble).
These network ensembles are  equivalent in the thermodynamic limit like the micro-canonical and the canonical ensemble of a gas of molecules. Nevertheless in  network theory it is often the  case that  we are interested in characterizing network ensembles with an  extensive number of constraints like the ones in which we fix the degree sequence (micro-canonical ensemble) or the expected degree sequence (canonical ensemble). In these cases we no longer observe the equivalence of the two conjugated  ensembles \cite{Kartik2009,Kartik2010}.
In the following section we will provide evidence that the configuration model of simplicial complexes and its conjugated canonical ensemble of simplicial complexes are not asymptotically equivalent.

\subsection{The entropy of the configuration model of simplicial complexes }

The entropy evaluates the logarithm of the (typical) number of simplicial complexes in the ensemble. This quantity characterizes the complexity of the constraints or in other words {\em how complex are the simplicial complexes in the ensemble} \cite{EPL2008}.  In fact an ensemble constructed from very stringent, complex constraints will give rise to few network realizations. Therefore the entropy of network ensembles can be used in a variety of inference problems \cite{Theta,Theta_jacopo}.
Moreover the entropy of conjugated ensembles can indicate whether the two ensembles are asymptotically equivalent.  In fact, if  the entropy of two conjugated ensembles is not the same in the large network limit, the two ensembles are not asymptotically equivalent. In this paragraph we provide a  summary of the results obtained for the entropy of the configuration model of simplicial complexes.The details of the derivations will be given in the Appendices.

The entropy $\Sigma$ of the micro-canonical ensemble defined by the configuration model of  simplicial complexes  with given sequence $\{k_r\}$ of generalized degree of the nodes is defined as the logarithm of the number of simplicial complexes belonging to the ensemble, i.e.
\bea
\Sigma =\ln{\cal N} = \ln \left[\sum_{G} \prod_{r} \delta\left({k_{r},k_{d,0}(r)}\right) \right]. 
\label{rel0}
\eea
In fact it can be  easily shown  that  
\bea
\Sigma=\ln{\cal N}=-\sum_{G} P(G) \ln P(G)
\eea
where $P(G)$ is given by Eq. $(\ref{Pgmicro})$.
To distinguish the entropy $\Sigma$ from  the entropy $S$ of the canonical ensemble defined in Eq. $(\ref{Sentropy})$, we call $\Sigma$ the Gibbs entropy and $S$ the Shannon entropy.
When considering the entropies $\Sigma$ and $S$ of conjugated microcanonical  (configuration model) and canonical ensemble, with $\overline{k_r}=k_r$, $\forall r$ we obtain (see Appendix $\ref{ssigma}$ for details)
\bea
\Sigma=S-\Omega,
\label{Sigmasomega}
\eea
where $\Omega$ is the entropy of large deviation, which is the the logarithm of the probability that in the canonical network model  with expected generalized degree sequence $\{\overline{k_r}\}$ (with $\overline{k_r}=k_r$), the generalized degrees of the nodes take exactly the values $k_{d,0}(r)=k_r$.
Therefore  $\Omega$ can  be expressed as 
\bea
\Omega=-\ln\left[\sum_{G} P(G)\prod_{r} \delta\left({k_{r},k_{d,0}(r)}\right)\right],
\label{O3}
\eea 
where
\bea
P(G)= \prod_{\alpha\in{\cal Q}_{d}(N)}\Big[p_{\alpha}^{a_{\alpha}} (1- p_{\alpha})^{1-a_{\alpha}}\Big],
\eea
with
\bea
p_{\alpha}=\frac{e^{-\sum_{r \subset \alpha}\lambda_{r} }}{1 + e^{ -\sum_{r\subset \alpha}\lambda_{r}}},
\eea
and
\bea 
k_r=\sum_{\alpha|r\subset \alpha}p_{\alpha}.
\eea
Since $\Omega$ is non-negative, Eq. $(\ref{Sigmasomega})$ shows that the Gibbs entropy $\Sigma$ is less than or equal to the Shannon entropy $S$ and when $\Omega$ is not negligible, the two entropies  are not the same, indicating a non equivalence of the micro-canonical (configuration model) and the canonical ensemble for simplicial complexes.

For simplicial complexes with the structural cutoff, $\Omega$ given by Eq. $(\ref{O3})$ can be calculated using the saddle point approximation (see Appendix $\ref{somega}$ for details) obtaining
\bea
\Omega=-\sum_{r=1}^N \ln \left[\pi_{k_r}(k_r)\right]
\label{Omegasol}
\eea
where $\pi_{k_r}(k_r)$ is the Poisson distribution with average $k_r$ evaluated at $k_r$, i.e.
\bea
\pi_{k_r}(k_r)=\frac{1}{k_r!}k_r^{k_r}e^{-k_r}.
\eea
This expression is easily interpreted. In fact in the canonical ensemble the generalized degree of each node follows a Poisson distribution with average $\overline{k_r}=k_r$.
Therefore the probability that each of these generalized degrees takes exactly the value $k_{d,0}(r)=k_r$ is given by $\pi_{k_r}(k_r)$.
We note here that $\Omega$ given by Eq. $(\ref{Omegasol})$ is an extensive quantity and therefore the Gibbs entropy $\Sigma$ is significantly different from the Shannon entropy $S$ of the canonical ensemble, implying that the two conjugated ensembles are not asymptotically equivalent.

\subsection{The asymptotic formula for the number of simplicial complexes in the configuration model with structural cutoff}

The Gibbs entropy $\Sigma$   can be evaluated using Eq. $(\ref{Sigmasomega})$ together with Eq. (\ref{Omegasol})   and the expression given by Eq. $(\ref{S2})$ for the Shannon entropy, as long as we are in  presence of the structural cutoff $K_d$ defined in Eq. $(\ref{Kd2})$), getting
\bea
\Sigma&=&-\sum_{\alpha\in{\cal Q}_{d}(N)}\left[p_{\alpha}\ln p_{\alpha}+(1-p_{\alpha})\ln(1-p_{\alpha})\right]\nonumber \\
&&+\sum_{r=1}^N \ln \frac{k_r^{k_r}e^{-k_r}}{k_r!},
\label{scom}
\eea
where in presence of the structural cutoff the probabilities $p_{\alpha}$ are given by Eq. ($\ref{pa}$) with $\overline{k_r}=k_r$ for every node $r$.
Substituting the expression for $p_{\alpha}$ into Eq. $(\ref{scom})$ we get the asymptotic expression for the logarithm of the number of simplicial complexes  ${\cal N}$ in the configuration model
\bea
\Sigma&=&\ln{\cal N}\nonumber\\
&=&\frac{d}{d+1}\ln (\avg{k}N)!-\sum_{r=1}^N \ln k_r ! -\frac{\avg{k}N}{d+1}\ln d!\nonumber \\
&&-\frac{d!}{2(d+1)(\avg{k}N)^{d-1}}\left(\frac{\avg{k^2}}{\avg{k}}\right)^{d+1}.
\eea
Therefore, the asymptotic expression for the number ${\cal N}$ of simplicial complexes in the configuration model is given by 
\bea
{\cal N}&=&\frac{[(\avg{k}N)!]^{d/(d+1)}}{\prod_{r=1}^{N}k_r!}\frac{1}{(d!)^{\avg{k}N/(d+1)}}\nonumber \\
&&\hspace{-16mm}\times\exp\left[-\frac{d!}{2(d+1)(\avg{k}N)^{d-1}}\left(\frac{\avg{k^2}}{\avg{k}}\right)^{d+1}+{\cal O}(\ln N)\right].
\label{CB}
\eea
This expression is the generalization of the Canfield-Bender formula \cite{CBender} for the ensemble of networks with given degree sequence. In fact for $d=1$ it is reduced to the Canfield-Bender formula.
Interestingly we observe that the asymptotic number ${\cal N}$ of simplicial complexes in the configuration model depends on the distribution of the generalized degrees of the nodes and that this dependency remains important even for generalized degree sequences with the same average $\langle k \rangle$. This shows that the complexity of the ensemble  depends strongly on the statistical characteristics of the generalized degree sequence.
As observed in Ref. \cite{EPL2008} in the context of simple networks,  it can also be shown for simplicial complexes of dimension $d>1$ that scale-free distributions of generalized degrees with the same average $\avg{k}$ but with decreasing power-law exponent $\gamma$ correspond to more complex ensembles of simplicial complexes. In fact they are  characterized by a smaller entropy $\Sigma$ and a smaller asymptotic number ${\cal N}$ of simplicial complexes.

\subsection{Combinatorial arguments for Eq. $(\ref{CB})$}
\label{com}
\begin{figure}
\begin{center}
{\includegraphics[width=0.99\columnwidth]{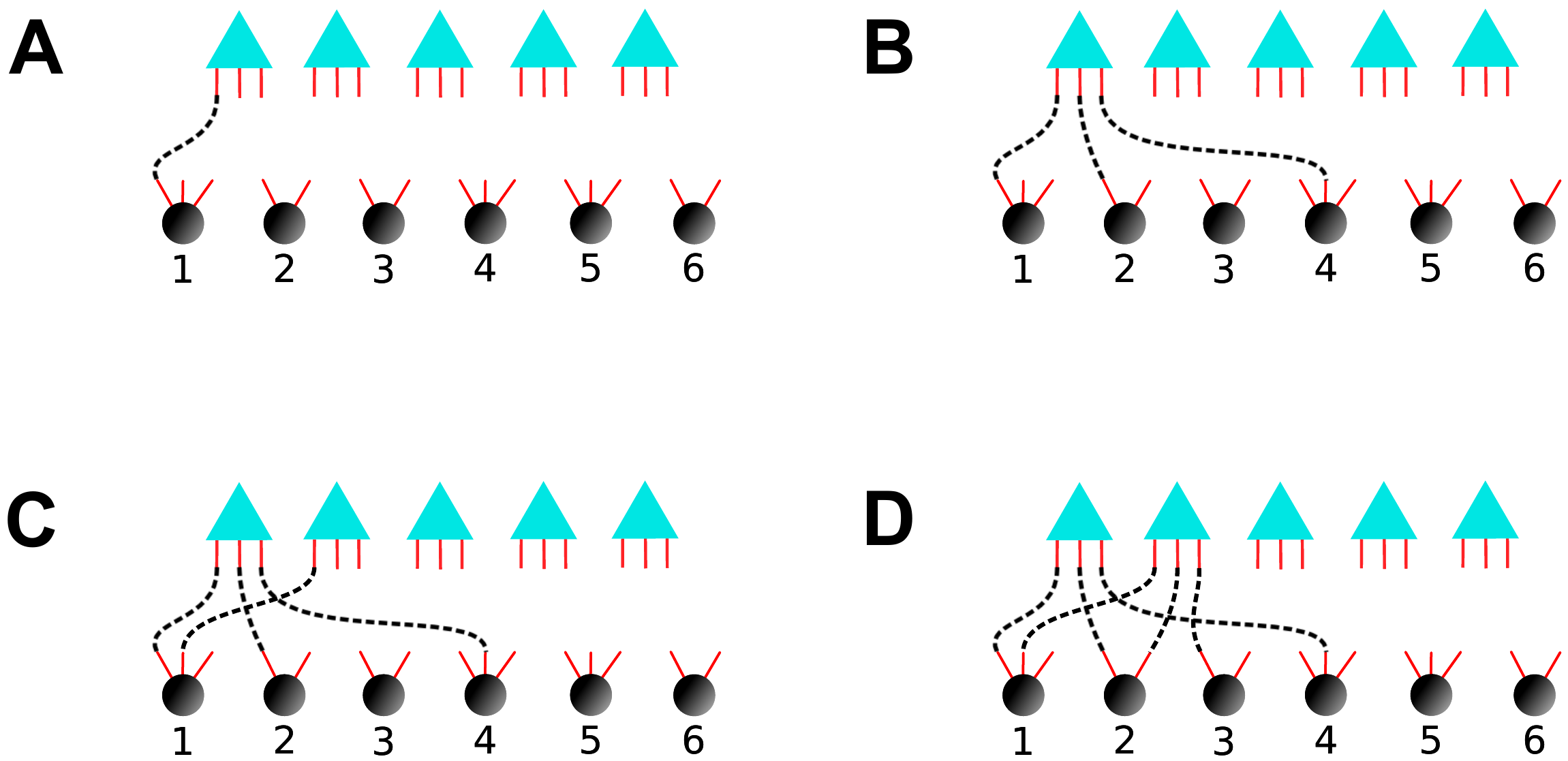}}
\end{center} 
\caption{(Color online) The subsequent matching of the node stubs with the stubs of the factor nodes is shown here  in order to  justify Eq. $(\ref{match})$ evaluating the asymptotic number of matchings in the absence of forbidden moves. First a random stub of a node is matched with a stub of an arbitrary unmatched factor node (panel A).  Subsequently all the remaining $d$ stubs of the factor node are matched with stubs of the nodes (panel B). At this point the unmatched factor nodes are reduced by one. An unmatched node-stub  is matched  to a stub of an arbitrary unmatched factor node (panel C). Subsequently  all the remaining stubs of the second factor node are matched with stubs of the nodes (panel D). This procedure continues in the absence of forbidden moves, until all of the stubs of the nodes are matched with all the stubs of the factor nodes. By calculating the probability of these moves, as describe in Sec. $\ref{com}$ we can derive Eq. $(\ref{match})$. }
\label{fig:comb}
\end{figure}

The asymptotic combinatorial expression (Eq. $(\ref{CB})$) can be explained using combinatorial arguments, similar to the ones used to explain the Canfield-Bender formula in Ref. \cite{PREGB_2009}.
In fact the factor 
\bea
\frac{[(\avg{k}N)!]^{d/(d+1)}}{\prod_{r=1}^{N}k_r!}\frac{1}{(d!)^{\avg{k}N/(d+1)}}
\label{match}
\eea
 counts all the possible combinations of the stubs of the nodes in groups of $d+1$ stubs when we disregard forbidden moves. 
In other words Eq. $(\ref{match})$ counts all the possible matchings between the stubs of the nodes and the stubs of the factor nodes obtained by following the algorithm described in Sec . $\ref{conf_al}$, considering the fact that the factor nodes are not labelled and neglecting the occurrence of forbidden moves.
In fact, if we want  to construct a simplicial complex with a given sequence of the  generalized degree of the nodes $\{k_r\}$, the first step is to take a stub of a node and match it with a stub of an unmatched factor node. Since the factor nodes are not labelled  every unmatched factor node is equivalent  and therefore there  is a unique way to match a given stub of the node with an unmatched unlabelled factor node.
Subsequently we proceed  with  matching the remaining $d$ stubs of this newly matched factor node. In order to do this, we  choose an unordered set of $d$ of the remaining stubs of the nodes. We have $[\avg{k}N-1][\avg{k}N-2]\ldots [\avg{k}N-(d+1)-1]/d!$ ways to perform this step if we neglect forbidden moves.
Once these stubs have been chosen, all the stubs of the  first factor node are now matched to $d+1$ stubs of the nodes. The remaining factor nodes are all unmatched. 
The next step is to take one of the remaining $\avg{k}N-(d+1)$ stubs of the nodes and to match it to an arbitrary unmatched factor node. Since the unmatched factor nodes are all equivalent, there is a unique way to to this. Subsequently we proceed as we have done previously and we find $d$ remaining  stubs of the nodes to be matched to the second factor node.  There are $\avg{k}N-(d+2)$ remaining stubs to choose from so there are $[\avg{k}N-(d+2)][\avg{k}N-(d+3)]\ldots [\avg{k}N-2(d+1)-1]/d!$ ways of selecting this unordered set of $d$ stubs. For a visual representation of these steps see Figure $\ref{fig:comb}$.
It is easy to see in this way that we can proceed by matching all the stubs of the nodes with the indistinguishable factor nodes in 
\bea
\frac{(\avg{k}N-1)!}{\prod_{s=1}^{\frac{\avg{k}N}{d+1}-1}[\avg{k}N-s(d+1)])}\frac{1}{{(d!)^{\avg{k}N/(d+1)}}}
\label{match3}
\eea
ways since $\avg{k}N/(d+1)=M$ counts the number of factor nodes. 
As long as $d$ is finite and $N\gg1$, we can use the following approximation, 
\bea
&&\frac{(\avg{k}N-1)!}{\prod_{s=1}^{\frac{\avg{k}N}{d+1}-1}[\avg{k}N-s(d+1)])}\simeq [(\avg{k}N)!]^{\frac{d}{d+1}},
\eea
getting the asymptotic approximation for Eq.$(\ref{match3})$ given by 
\bea
\frac{[(\avg{k}N)!]^{d/(d+1)}}{(d!)^{\avg{k}N/(d+1)}}.
\label{match2}
\eea
In order to find the number of distinct matchings given by Eq. $(\ref{match})$ we need to observe that all the permutations of the stubs of each single  node give equivalent matchings. We need therefore to divide the expression found in Eq. $(\ref{match2})$ by $\prod_{r=1,N}k_r!$ getting Eq. $(\ref{match})$.
Finally, the exponential term in Eq. $(\ref{CB})$ needs to be interpreted as the term that corrects for the forbidden matchings.
\section{Natural correlations of the configuration model of  simplicial complexes}

\begin{figure*}
\begin{center}
\includegraphics[width=1.99\columnwidth]{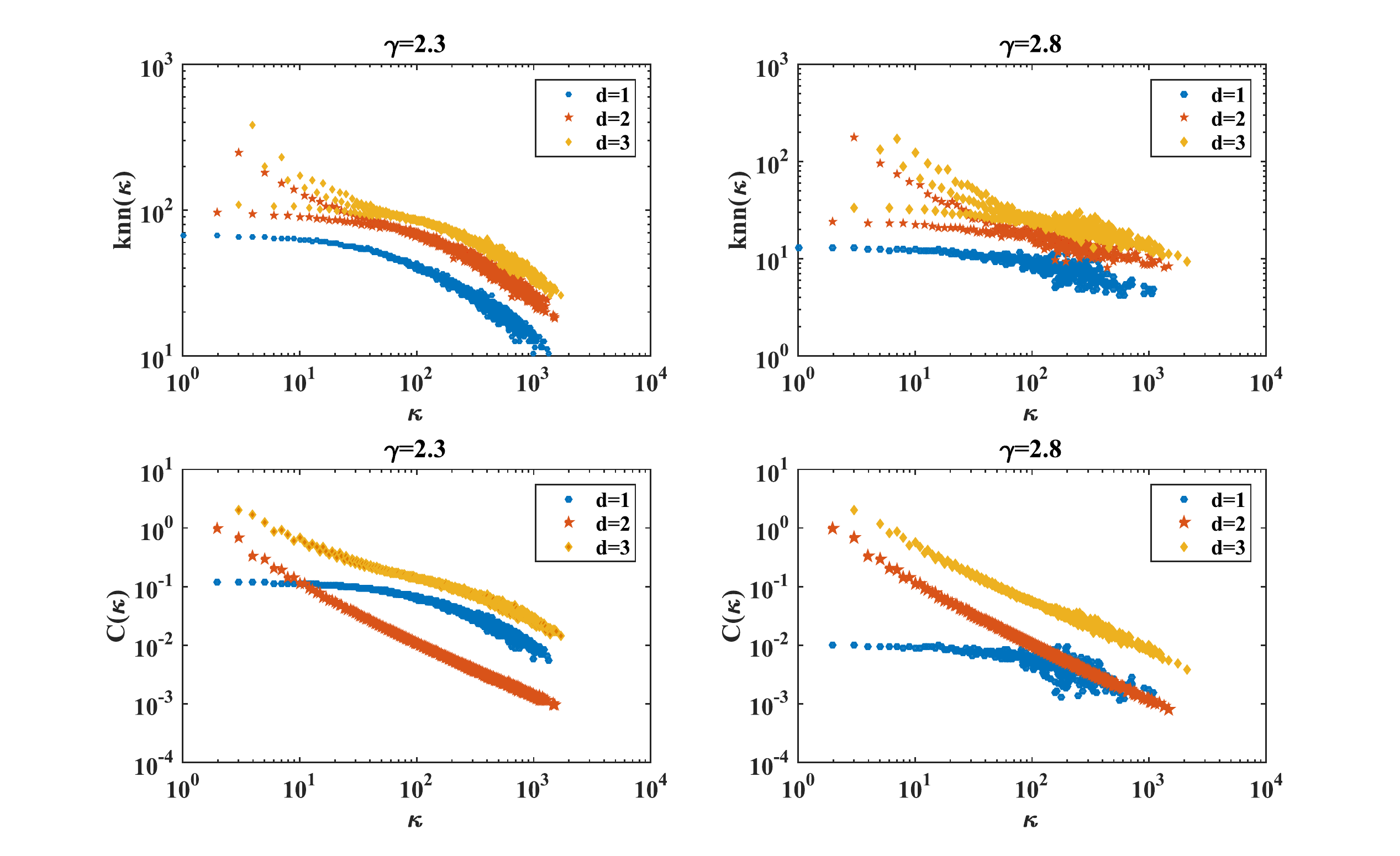}
\end{center}
\caption{(Color online) The average degree $knn(\kappa)$ of the neighbors of the nodes of degree $\kappa$ and the average clustering coefficient $C(\kappa)$ of the nodes of degree $\kappa$, for simplicial complexes of dimension $d=1,2,3$ constructed according to the configuration model with distribution of the generalized degrees of the nodes $P_{d,0}(k)$ given by Eq. $(\ref{Sf})$ and $\gamma=2.3,2.8$. The simplicial complexes have $N=10^4$ nodes, and $n_F=70$. The data are averaged over $100$ realizations.}
\label{fig4}
\end{figure*}

The configuration model of simplicial complexes without structural cutoff develops significant degree correlations. 
In order to characterize the degree correlations present in the simplicial complexes of different dimension $d$, we have consider the simplicial complexes constructed by the configuration model with  scale-free distribution $P_{d,0}(k)$ of the generalized degree of the nodes $k_{d,0}=k$. 
The distribution $P_{d,0}(k)$ of the generalized degree of the nodes is given by 
\bea
P_{d,0}(k)=Ck^{-\gamma},
\label{Sf}
\eea
with minimal generalized degree $m=1$.
The generated simplicial complexes can be analyzed by means of the well established tools used in network theory. In particular a network structure can be extracted from the simplicial complexes by assuming that two nodes are linked if they belong at least to a common simplex.
We will call the adjacency matrix of this network ${\bf \hat{a}}$ and the degree of the generic node $r$  in this network $\kappa_r$. 
The correlations existing in this network can be characterized  by means of the average degree $knn(\kappa)$ of the neighbors of the nodes of degree $\kappa$, and the average clustering coefficient $C(\kappa)$ of the nodes of degree $\kappa$.
These functions are plotted in Figure $\ref{fig4}$ for simplicial complexes with generalized degree distribution $P_{d,0}(k)$ given by Eq. $(\ref{Sf})$ and $\gamma=2.3,2.8$.
These results show that natural degree correlations occur in these models.
The average clustering coefficient $C(\kappa)$ increases with the increasing dimensionality $d$ of the simplicial complex, and the shape of the function $C(\kappa)$ also in strongly dependent on the dimensionality $d$.
On the contrary,   $knn(\kappa)$ does not appear to change so dramatically with the dimensionality $d$ of the simplicial complex.

\section{Conclusions}

Simplicial complexes  of dimension $d>1$ encode information about  interactions occurring between more than two nodes while simplicial complexes of dimension $d=1$ are simple networks describing only pairwise interactions. As such, simplicial complexes are a generalization of network structures that can be extremely useful for analyzing a large variety of complex interacting systems ranging from brain networks to social networks.
As   novel approaches to data analysis of networked systems require the characterization of complex datasets in terms of  simplicial complexes, building null models for these structures is increasingly important for the advance of the field.
Here we have characterized the structural properties of simplicial complexes using the generalized degrees, which  capture fundamental properties of their $\delta$-faces. We have fully investigated the configuration model for simplicial complexes with statistical mechanics techniques  relating its properties  with the ones of the conjugated canonical ensemble of simplicial complexes (also called  the exponential random simplicial complex).
The entropy of these ensembles is derived here with analytical techniques, opening  the possibility to use this quantity as an information theory measure for inference problems on simplicial complexes.
Additionally we have found an expression for the structural cutoff of simplicial complexes that generalizes the structural cutoff of the configuration model of simple networks.
Finally we have provided algorithms for generating simplicial complexes belonging to the configuration model and the canonical ensembles studied in this paper, and we have numerically investigated the natural correlations emerging in these models. 

In conclusion we believe that this paper provides a full account of two of  the most fundamental  equilibrium models of simplicial complexes which can be used as null models for investigating the structure of simplicial complexes, or  for studying dynamical processes. We believe that these models constitute only the first step in modelling simplicial complexes with equilibrium statistical mechanics tools and that our work will open new perspectives for investigating  a new generation of  equilibrium models for simplicial complexes.

\appendix

\section{Derivation of Eq. ($\ref{Sigmasomega}$) relating  the entropy $\Sigma$ configuration model and the entropy $S$ of the canonical ensemble}
\label{ssigma}
In this section we want to derive the relation
\bea
\Sigma=S-\Omega,
\label{rel1}
\eea
where $\Sigma$ is the entropy of the configuration model of simplicial complexes given by the logarithm of the number of graphs satisfying hard constraints on the generalized degree $k_{d,0}(r)=k_r$ for vertices $r=1,...,N$, i.e.
\bea
\Sigma = \ln \left[\sum_{G} \prod_{r} \delta\left({k_{r},k_{d,0}(r)}\right)\right], \eea
$S$ is the entropy of the canonical ensemble of simplicial complexes enforcing the expected generalized degree given by $\overline{k_{d,0}(r)}=k_r$
and $\Omega$ is the entropy of large deviation which is the logarithm of the probability that in the canonical ensemble mentioned above, the generalized degree $k_{d,0}(r)$ take exactly the values $k_{d,0}(r)=k_r$.
The entropy $S$ is given by
 
\bea
S&=&-\sum_{\alpha}\left[p_{\alpha}\ln p_{\alpha}+(1-p_{\alpha})\ln(1-p_{\alpha})\right]
\label{Shan}
\eea

while $\Omega$ is given by
\bea
\Omega=-\ln\left[\sum_{G} P(G)\prod_{r} \delta\left({k_{r},k_{d,0}(r)}\right)\right]
\label{Omega2}
\eea
In order to derive Eq. $(\ref{rel1})$ we use the integral representation of the Kroenecker delta
\bea
\delta (x , y) = \int_{-\pi}^{\pi} \frac{d\omega}{2\pi} e^{i\omega x - i\omega y}, \eea
obtaining for $\Sigma$
\bea \Sigma =  \ln \left[\sum_{G} \prod_{r} \int_{-\pi}^{\pi} \frac{d\omega_{r}}{2\pi} e^{i\omega_r k_r - i\omega_r \sum_{\alpha|r\subset\alpha} a_{\alpha}} \right] \eea
Summing over all simplicial complexes $G$ is equivalent to summing over all adjacency tensors $\bf a$ of elements $a_{\alpha}=0,1$. Performing the sum we get
\bea
 \Sigma &=& \ln \int_{-\pi}^{\pi} \left(\prod_{r} \frac{d\omega_{r}}{2\pi}\right) e^{i \sum_{r} \omega_r k_r} \prod_{\alpha\in{\cal Q}_d(N)} \Big[1 + e^{ - i \sum_{r\in \alpha}\omega_r  }\Big]\nonumber \\
&=&\ln \int_{-\pi}^{\pi} \left(\prod_{r} \frac{d\omega_{r}}{2\pi}\right) e^{F({\boldsymbol{\omega}}, \bf{k})},  
\label{int}\eea
where the function $F({\boldsymbol{\omega}}, {\bf{k}})$ is given by 
\bea
 F({\boldsymbol{\omega}}, {\bf{k}}) = i\sum_{r} \omega_r k_r  + \sum_{\alpha\in {\cal Q}_d(N)}\ln \left[1+ e^{ -i \sum_{r\in \alpha} \omega_r } \right].   
\eea

We now apply a change of variables $\omega_r \to z_r$ where 
\bea
\omega_r = z_r + \omega_{r}^{\star},
\label{shift}
\eea
and where $\{\omega_{r}^{\star}\}$ indicates the solution to   the equation
\bea
\frac{d}{d\omega_r}F({\boldsymbol{\omega}}, {\bf{k}})= 0.
\eea
We note, that the above equations imply  that the $\omega_{r}^{\star}$ are related to our choice of generalized degree sequence by
\bea
k_r = \sum_{\alpha | r\subset \alpha} \frac{e^{-i\sum_{m\subset \alpha}\omega_m^{\star} }}{1 + e^{-i\sum_{m\subset \alpha}\omega_m^{\star}}}. 
\label{krc}
\eea
We can therefore  identify $i\omega_{r}^{\star}$ with the parameters $\lambda_r$ of the canonical ensemble in which we enforce that the expected generalized degree of the nodes $\overline{k_{d,0}(r)}$ take the same value of the hard constrained generalized degrees, 
i.e.
\bea
\overline{k_{d,0}(r)}=k_r.
\eea
In fact by setting $\lambda_r =i\omega_{r}^{\star}$  we get
\bea
p_{\alpha}&=& \frac{e^{-\sum_{m\subset \alpha}\lambda_m}}{1 + e^{-\sum_{m\subset \alpha}\lambda_m}}\nonumber \\
&= &\frac{e^{-i\sum_{m\subset \alpha}\omega_m^{\star}}}{1 + e^{-i\sum_{m\subset \alpha}\omega_m^{\star}}} 
\label{pa2}
\eea
and Eq. $(\ref{krc})$ reads 
\bea
 k_{r} = \overline{k_{d,0}(r)}= \sum_{\alpha|r\subset \alpha} p_{\alpha} .
\eea

Writing $F({\boldsymbol{\omega}}, {\bf{k}})$ in terms of our new variable $z_r$ given by Eq. $(\ref{shift})$
we obtain
\bea
 F({\boldsymbol{z}},{\boldsymbol{\omega^{\star}}},  {\bf{k}}) &=& i \sum_r \omega_{r}^{\star}k_r    +\sum_{\alpha\in {\cal Q}_d(N)}\ln\Bigg[1+ e^{ -i \sum_{r\in \alpha} \omega_r^{\star}} \Bigg]   \nonumber\\
&&\hspace*{-25mm}+ i \sum_r z_{r}k_r + \sum_{\alpha\in {\cal Q}_d(N)}\ln\Bigg[1-p_{\alpha}+ p_{\alpha}e^{-i \sum_{r\in \alpha} z_r } \Bigg],  
\label{FA}
 \eea
where $p_{\alpha}$ is given by Eq. $(\ref{pa2})$.
We identify the first two terms of this expression with the entropy of the canonical ensemble $S$ given by Eq. (\ref{S2}). In fact 
\bea
S&=&-\sum_{\alpha}\left[p_{\alpha}\ln p_{\alpha}+(1-p_{\alpha})\ln(1-p_{\alpha})\right]\nonumber \\
\hspace{-5mm}&\hspace{-7mm}=&\hspace{-5mm} i \sum_r \omega_{r}^{\star}k_r    +\sum_{\alpha\in {\cal Q}_d(N)}\ln\Bigg[1+ e^{ -i \sum_{r\in \alpha} \omega_r^{\star}} \Bigg], 
\label{SA}
\eea
where in the last expression we have substituted in the expression for $p_{\alpha}$ given by Eq. $(\ref{pa2})$.
Finally, using Eq. $(\ref{SA})$ and inserting the expression found for $F({\boldsymbol{z}},{\boldsymbol{\omega^{\star}}},  {\bf{k}})$ (Eq. $(\ref{FA})$) back in to Eq. $(\ref{int})$ we obtain

\bea \Sigma &=& \ln \left\{e^{S}\int_{-\pi}^{\pi} \left(\prod_{r} \frac{dz_{r}}{2\pi}\right) e^{i \sum_{r} z_r k_r} \right.\nonumber \\
&&\left.\times \prod_{\alpha\in{\cal Q}_d(N)} \Big[1-p_{\alpha} +p_{\alpha} e^{ - i \sum_{r\in \alpha}z_r  }\Big]\right\}\nonumber \\
&&\hspace*{-16mm}=\ln \left[e^{S}\sum_{\{a_{\alpha}\}}\prod_{\alpha}p_{\alpha}^{a_{\alpha}}(1-p_{\alpha})^{1-a_{\alpha}}\prod_{r} \delta\left({k_{r}, \sum_{\alpha'|r\subset \alpha'}a_{\alpha'}}\right)\right]\nonumber
\eea
Therefore $\Sigma$ can be written as
\bea
\Sigma= \ln \left[e^{S} \sum_{G} P(G) \prod_{r} \delta\left({k_{r},\sum_{\alpha| r\subset \alpha}a_{\alpha}}\right)\right],
\eea
where $P(G)$ is given by Eq. ($\ref{PG}$).
Using the definition of $\Omega$ given by Eq. $(\ref{Omega2})$ it follows that 
\bea
\Sigma= S -\Omega.
\eea

\section{Derivation of the Eq. ($\ref{Omegasol}$) for $\Omega$}
\label{somega}
In this appendix we derive Eq. $(\ref{Omegasol})$ for $\Omega$ in the presence of the  structural cutoff. The quantity $\Omega$ indicates the logarithm of the probability that in a canonical ensemble of simplicial complexes enforcing the sequence of expected degree of the nodes $\{\overline{k_r}=k_r\}$ we observe a simplicial complex realization in which the sequence of the generalized degree of the nodes is exactly $\{k_r\}$. By indicating with $P(G)$ the probability of a simplicial complex in the canonical ensemble, we have 
\bea
\Omega &&= -\ln \sum_{G} P(G) \prod_{r} \delta\left({k_{r}, k_{d,0}(r)}\right)\nonumber \\
\nonumber \\
&&\hspace{-7mm}=-\ln \sum_{G}\prod_{\alpha}  p_{\alpha}^{a_{\alpha}}(1-p_{\alpha})^{1-a_{\alpha}}\prod_{r}\delta\left({k_{r}, k_{2,0}(r)}\right)
\eea
where, in presence of the structural cutoff, the probabilities $p_{\alpha}$ are given by Eq. $(\ref{pa2})$.
In order to evaluate $\Omega$, we  use the integral representation of the Kroenecker delta
\bea
\delta (x , y) = \int_{-\pi}^{\pi} \frac{d\omega}{2\pi} e^{i\omega x - i\omega y} ,
\eea
getting
\bea
\Omega&&= -\ln \sum_{G}\prod_{\alpha}  p_{\alpha}^{a_{\alpha}}(1-p_{\alpha})^{1-a_{\alpha}}\nonumber \\
&&\times\prod_{r} \int_{-\pi}^{\pi} \frac{d\omega_{r}}{2\pi} e^{i\omega_{r} k_{r} - i\omega_{r}\sum_{\alpha'|r\subset \alpha'} a_{\alpha'}}\nonumber \\
&&=-\ln \int_{-\pi}^{\pi} \prod_{r} \frac{d\omega_r}{2\pi}  e^{{\cal G}[\{\omega_r\}]}
\eea
where 
\bea
{\cal G}[\{\omega_r\}]=&&{i\sum_i\omega_{r} k_{r}+\sum_{\alpha}\ln\left[1+p_{\alpha} \left(e^{-i \sum_{r\subset \alpha}\omega_r}-1\right)\right]}.\nonumber
\eea
For an uncorrelated simplicial complex ensemble with structural cutoff and with $p_{\alpha}$ given by Eq. $(\ref{pa})$ and  $p_{\alpha}\ll1$ we can approximate  ${\cal G}[\{\omega_r\}]$  as
\bea
{\cal G}[\{\omega_r\}]={i\sum_r\omega_{r} k_{r}+\sum_{\alpha}p_{\alpha} \left(e^{-i \sum_{r\subset \alpha}\omega_r}-1\right)}.
\eea
Using the explicit factorized expression for $p_{\alpha}$ in the presence of the structural cutoff ( Eq. $(\ref{pa})$), we observe that we can write
\bea
\hspace*{-4mm}\sum_{\alpha}p_{\alpha} \left(e^{-i \sum_{r\subset \alpha}\omega_r}-1\right)=\frac{d!}{(d+1)!}\avg{k}N\left(\nu^{d+1}-1\right)
\eea
where
\bea
\nu=\sum_r \frac{k_r}{\avg{k}N}e^{-i\omega_r}.
\label{alpha}
\eea
We now introduce the density
\bea
c(\omega|  k)=\frac{1}{N_{k}}\sum_{r} \delta(\omega - \omega_{r})\delta\left({k , k_{r}}\right)
\eea
where 
\bea
N_{k} = N P_{d,0}(k)
\eea
indicates the number of nodes with generalized degree of the nodes $k_r=k$ and $P_{d,0}(k)$ indicates the distribution of the generalized degree of the nodes.
We can therefore express $\nu$ given by Eq. $(\ref{alpha})$ in terms of $c(\omega|k)$ obtaining
\bea
\nu=\sum_k \frac{k}{\avg{k}}P_{d,0}(k)\int d\omega e^{-i\omega}c(\omega|k).
\label{alpha2}
\eea

Using the delta functions 
\bea
&&\delta (c(\omega| k) , \frac{1}{N_k}\sum_{r} \delta\left[\omega - \omega_{r})\delta\left({k , k_{r}}\right)\right] \nonumber \\
&&\hspace{-8mm}= \int_{-\pi}^{\pi} \frac{d\hat{c}(\omega|k)}{2\pi N_k} e^{i\hat{c}(\omega|k)[N_k c(\omega|k) - \sum_{r} \delta(\omega - \omega_{r}) \delta\left({k , k_{r}}\right)]} 
\eea
we can now express $\Omega$ as
\bea
\Omega = -\ln \int {\cal D} c(\omega| \hat{\lambda},k) {\cal D} \hat{c}(\omega| \hat{\lambda}, k) e^{NF[c(\omega|  k),\hat{c}(\omega| k) ]},\nonumber
\label{O2}
\eea
where $F[c(\omega|  k),\hat{c}(\omega| k) ]$ is given by 
\bea
&&{F[c(\omega|  k),\hat{c}(\omega| k) ]}=i\sum_k P_{d,0}(k)\int d\omega\hat{c}(\omega|k) c(\omega|k)\nonumber \\
&&\hspace*{-4mm}+\frac{d!}{(d+1)!}\avg{k}\left(\nu^{d+1}-1\right)\nonumber \\
&&+\sum_k P_{d,0}(k)\ln \int \frac{d\omega}{2\pi}e^{i\omega k-i\hat{c}(\omega|k)}.
\label{Fc}
\eea
We evaluate the integral  $(\ref{O2})$ with the saddle point method. The saddle point equations read,
\bea
\frac{\partial F[c(\omega| k),\hat{c}(\omega|  k) ]}{\partial c(\omega| k)}&=&0,\nonumber\\
\frac{\partial F[c(\omega|  k),\hat{c}(\omega|  k) ]}{\partial \hat{c}(\omega| k)}&=&0.\nonumber 
\eea
Which gives us
\bea
-i\hat{c}(\omega| k) &=& k \nu^d e^{-i\omega},\label{saddle1} \\
c(\omega| k) &=& \frac{\frac{1}{2\pi} e^{i\omega k - i \hat{c}(\omega|  k)}}{\int \frac{d\omega}{2\pi}e^{i\omega k - i \hat{c}(\omega| k)}}. \label{saddle2}
\eea
Using Eq. (\ref{saddle1}) we observe that the integral appearing in Eq. $(\ref{saddle2})$ can be expressed in terms of $\nu$, obtaining 
\bea
&&\int \frac{d\omega}{2\pi}e^{i\omega k - i \hat{c}(\omega| \hat{\lambda}, k)}=\int \frac{d\omega}{2\pi}e^{i\omega k+k\nu^d e^{-i\omega}}\nonumber \\
&&=\int \frac{d\omega}{2\pi}e^{i\omega k}\sum_h(\nu^d k)^h e^{-i\omega h}\frac{1}{h!}=\frac{(\nu^dk)^k}{k!}.
\label{Ic}
\eea 
Substituting this result in to Eq. $(\ref{saddle2})$, we get
\bea
c(\omega|k)&=&\frac{k!}{2\pi (\nu^d k)^k} e^{i\omega k - i \hat{c}(\omega|  k)},\nonumber \\
&&\frac{k!}{2\pi (\nu^d k)^k} e^{i\omega k + k \nu^d e^{-i\omega}}.
\eea
Finally, we can substitute this expression in to the definition of $\nu$ given by Eq. $(\ref{alpha2})$ obtaining
\bea
\nu&=&\sum_k \frac{k}{\avg{k}}P_{d,0}(k)\int d\omega e^{-i\omega}c(\omega|k)\nonumber\\
&=&\sum_k \frac{k}{\avg{k}} P_{d,0}(k)\frac{k!}{(\nu^dk)^k}\int \frac{d\omega}{2\pi}e^{i\omega (k-1)+k\nu^d e^{-i\omega}}\nonumber \\
&=&\sum_k \frac{k}{\avg{k}}P_{d,0}(k)\frac{k!}{(\nu^dk)^k}\int \frac{d\omega}{2\pi}e^{i\omega (k-1)}\sum_h (\nu^d k)^h e^{-i\omega h}\frac{1}{h!}\nonumber \\
&=&\sum_k \frac{k}{\avg{k}} P_{d,0}(k) \frac{k!}{(\nu^dk)^k}\frac{(\nu^dk)^{k-1}}{(k-1)!}=\nu^{-d}.
\label{alphac}
\eea
Therefore, $\nu$ is the solution of  the equation $\nu=\nu^{-d}$, and so we have
\bea
\nu=1.
\eea
Using this result, and Eq. (\ref{Ic}) it is immediate to show that the value of the functional ${F[c(\omega|  k),\hat{c}(\omega| k) ]}$ (Eq. (\ref{Fc})) at the saddle point is given by 
\bea
{F[c(\omega|  k),\hat{c}(\omega| k) ]}&=&i\sum_k P_{d,0}(k)\int d\omega\hat{c}(\omega|k) c(\omega|k)\nonumber \\
&&+\sum_k P_{d,0}(k) \ln\left[\frac{k^k}{k!}\right].
\eea
Proceeding as in Eq. $(\ref{alphac})$ it can be easily shown that 
\bea
&&i\sum_k P_{d,0}(k)\int d\omega\hat{c}(\omega|k) c(\omega|k)=\nonumber\\
\hspace*{3mm}&=&-\sum_k P_{d,0}(k) k \frac{k!}{k^k}\int d\omega {2\pi}e^{i\omega (k-1)+k e^{-i\omega}}\nonumber \\
\hspace*{3mm}&=&-\avg{k}=-\sum_k P_{d,0}(k) k.
\eea
Finally, evaluating the integral $(\ref{O2}$) at the saddle point we obtain the simple expression for $\Omega$ given by 
\bea
\Omega& =&- \ln\Bigg[N \sum_{k}P_{d,0}(k)\ln\Bigg({\frac{{k}^{k}}{k !}e^{-k}}\Bigg)\Bigg]\nonumber \\
&=& -\ln\Bigg[\sum_{r} \ln\Big(\pi_{k_{r}}(k_{r})\Big)\Bigg],
\eea
where $\pi_{k_r}(k_r)$ indicated the Poisson distribution with average $k_r$ calculated at $k_r$, i.e.
\bea
\pi_{k_r}(k_r)=\frac{{k_r}^{k_r}}{k_r !}e^{-k_r}.
\eea
\clearpage

\onecolumngrid

 \section*{SUPPLEMENTARY MATERIAL}
\subsection*{Introduction}
Simplicial complexes are a generalization of  simple networks  including interactions between more than two nodes (for example the interaction of three nodes in a triangle or four nodes in a tetrahedron). In this supplementary material we provide code extending the network configuration model (dimension $d=1$) to simplicial complexes of dimensions $d=2$ and $d=3$. This generalized configuration model generates simplicial complexes with given generalized degree sequence (the sequence of the number of triangles or tetrahedra incident to each node for $d=2$ and $d=3$ respectively). 
\subsection*{Codes for generating simplicial complexes using the configuration model}
\label{6}
In this section we provide three C codes for generating simplicial complexes in the configuration model in dimension $d=1,2,3$. The generalized degree sequences used in our codes are drawn randomly from a scale-free distribution, however the codes can easily be modified to take pre-specified generalized degree sequences as inputs. Additionally, alternative code can be found in the comments which may be used to draw the sequences from a Poisson distribution. A detailed description of the algorithm used in given in Sec. IVB of the main text. In the code there is an option to  allow the rejection of a    given maximum number  of  forbidden moves.
\subsubsection*{Code for $d=1$}


\begin{lstlisting}
/**************************************************************************************************
 * If you use this code, please cite G. Bianconi and O.T. Courtney
 * "Generalized network structures: the configuration model and the canonical ensemble of
 * simplicial complexes"
***************************************************************************************************
 * Code that  generates random simplicial complexes with scale-free generalized degree 
 * distribution.
 *
 * The option to use a Poisson distributed generalized degree distribution has also been included.  
 * The necessary code may be found in comments at the relevant points.
 *
 * This code uses:
 * N  Number of nodes in the simplicial complex
 * m  The minimum of the scale-free distribution
 * gamma2  Exponent of the scale-free distribution
 * lambda  Expected value of the Poisson distribution (commented-out)
 * Avoid  Whether or not 'back-tracking' is allowed when illegal matchings are proposed
 * (Avoid==1 allowed, Avoid==0 not allowed)
 * NX  Maximum number of 'back-tracks' before matching process restarts from an unmatched network
 *************************************************************************************************/

#include<stdio.h>
#include<stdlib.h>
#include<string.h>
#include<math.h>
#include<time.h>

#define N 10000
#define m 1
#define gamma2 2.3
/* #define lambda 10 */
#define Avoid 1
#define NX 80

int *kgi,*kg,***tri;

/*************************************************************************************************/
/* Randomly select an unmatched stub. Choose takes as its input a random number between 0 and the 
total number of stubs and gives as its output the index of the node of the selected stub */
int Choose(double x){
	int i1,i;
	for (i=0;i<N;i++){
		x-=kgi[i];
		if (x<0){
			i1=i;
			break;
		}
	}
	return(i1);
}
/*************************************************************************************************/

int main(int argc, char** argv){
	int i,j,nrun,j2,i1,i2,i3,naus,*knng,*pkg,*k,**l,*pk,*knn,n,**a,*Ck;
	double xaus, x;
	char filec[60];

	FILE *fp,*gp;

	gp=fopen("edge_list.txt","w");
	srand48(time(NULL));
	kgi=(int*)calloc(N,sizeof(int));
	kg=(int*)calloc(N,sizeof(int));
	k=(int*)calloc(N,sizeof(int));
	a=(int**)calloc(N,sizeof(int*));
	knng=(int*)calloc(N,sizeof(int));
	pkg=(int*)calloc(N,sizeof(int));
	knn=(int*)calloc(N,sizeof(int));
	pk=(int*)calloc(N,sizeof(int));
	Ck=(int*)calloc(N,sizeof(int));

	for(i=0;i<N;i++){
		a[i]=(int*)calloc(N,sizeof(int));
	}

	xaus=4;  

	while(xaus>2){
	/***********************************************************************************************/
	/* Initialization */
		for(i=0;i<N;i++){
		/* Nodes are assigned desired generalized degree according to a scale-free distribution */
			kgi[i]=(int)(m*pow(drand48(),-1./(gamma2-1.)));
			/* kgi[i]= poisson(lambda); */
			while(kgi[i]>(N-1)){
			/* Desired generalized degrees are re-drawn if they exceed the maximum possible generalized degree of a node (natural cut-off) */
				kgi[i]=(int)(m*pow(drand48(),-1./(gamma2-1.)));
				/* kgi[i]= poisson(lambda); */
			}
			kg[i]=0;  /* Generalized degree of node i intially set to 0 */
			k[i]=0;  /* Degree of node i intially set to 0 */
			for(j=0;j<N;j++){
				a[i][j]=0;
			}
		}
		xaus=0;
		for(i=0;i<N;i++){
			xaus+=kgi[i];
		}
		naus=0; /* Back-track counter initially set to zero */
	/***********************************************************************************************/
	/* Stubs matched */
		while((xaus>3)&&(naus<1+Avoid*NX)){
			/* Randomly select two nodes proportional to the number of unmatched stubs they have remaining. */
			x=xaus*drand48();
			i1=Choose(x);
			kg[i1]++;
			kgi[i1]--;
			xaus--;

			x=xaus*drand48();
			i2=Choose(x);
			kg[i2]++;
			kgi[i2]--;
			xaus--;

			/* Check proposed matching is legal */
			if((i1!=i2)&&(a[i1][i2]==0)){
				/* Proposed matching legal. Create link */
				a[i1][i2]=1;
				a[i2][i1]=1;
			}
			else{
				/* Proposed matching illegal. Back-track and increment back-track counter by one */
				naus++;
				if(Avoid==1){
					kg[i1]--;
					kgi[i1]++;
					kg[i2]--;
					kgi[i2]++;
				}
			}
		}
	}
/*************************************************************************************************/
/* Degrees calculated */
	for (i=0;i<N;i++){
		for(j=i+1;j<N;j++){
			if(a[i][j]>0){
				k[i]++;
				k[j]++;
			}
		}
	}
/*************************************************************************************************/
/* Print list of edges to file */
	for (i=0;i<N;i++){
		for(j=0;j<N;j++){
			if(a[i][j]==1){
				fprintf(gp,'%d %d\n',i,j)
			}
		}

	}
/*************************************************************************************************/
	fclose(gp);

	return 0;
}



\end{lstlisting}

\subsubsection*{Code for $d=2$}

\begin{lstlisting}
/**************************************************************************************************
 * If you use this code, please cite G. Bianconi and O.T. Courtney
 * "Generalized network structures: the configuration model and the canonical ensemble of
 * simplicial complexes"
***************************************************************************************************
 * Code that  generates random simplicial complexes with scale-free generalized degree 
 * distribution.
 *
 * The option to use a Poisson distributed generalized degree distribution has also been included.  
 * The necessary code may be found in comments at the relevant points.
 *
 * This code uses:
 * N  Number of nodes in the simplicial complex
 * m  The minimum of the scale-free distribution
 * gamma2  Exponent of the scale-free distribution
 * lambda  Expected value of the Poisson distribution (commented-out)
 * Avoid  Whether or not 'back-tracking' is allowed when illegal matchings are proposed
 * (Avoid==1 allowed, Avoid==0 not allowed)
 * NX  Maximum number of 'back-tracks' before matching process restarts from an unmatched network
 *************************************************************************************************/

#include<stdio.h>
#include<stdlib.h>
#include<string.h>
#include<math.h>
#include<time.h>

#define N 500
#define m 1
#define gamma2 2.5
/* #define lambda 10 */
#define Avoid 1
#define NX 15
#define figure 1


int *kgi,*kg,***tri;

/*************************************************************************************************/
/* Randomly select an unmatched stub. Choose takes as its input a random number between 0 and the 
total number of stubs and gives as its output the index of the node of the selected stub */
int Choose(double x){
	int i1,i;
	for (i=0;i<N;i++){
		x-=kgi[i];
		if (x<0){
			i1=i;
			break;
		}
	}
	return(i1);
}
/*************************************************************************************************/
/* Check if a triangle exists. Takes three nodes as an input and outputs 1 if there already exists a triangle incident to them and 0 otherwise. */
int Check(i1,i2,i3){
	int in,c=0;
	for(in=0;in<kg[i1]-1;in++){
		if(((tri[i1][0][in]==i2)&&(tri[i1][1][in]==i3))||((tri[i1][0][in]==i3)&&(tri[i1][1][in]==i2))){
			c=1;
			break;
		}
	}
	return(c);
}
/*************************************************************************************************/
/* Create triangle. Takes 3 nodes as an input and creates a triangle incident to them. */
void Triangle(int i1, int i2,int i3){
	int iaus;
	tri[i1][0]=(int*)realloc(tri[i1][0],kg[i1]*sizeof(int));
	tri[i1][1]=(int*)realloc(tri[i1][1],kg[i1]*sizeof(int));
	tri[i2][0]=(int*)realloc(tri[i2][0],kg[i2]*sizeof(int));
	tri[i2][1]=(int*)realloc(tri[i2][1],kg[i2]*sizeof(int));
	tri[i3][0]=(int*)realloc(tri[i3][0],kg[i3]*sizeof(int));
	tri[i3][1]=(int*)realloc(tri[i3][1],kg[i3]*sizeof(int));
	tri[i1][0][kg[i1]-1]=i2;
	tri[i1][1][kg[i1]-1]=i3;
	tri[i2][0][kg[i2]-1]=i1;
	tri[i2][1][kg[i2]-1]=i3;
	tri[i3][0][kg[i3]-1]=i1;
	tri[i3][1][kg[i3]-1]=i2;
}
/*************************************************************************************************/

int main(int argc, char** argv){
	int i,j,nrun,j2,i1,i2,i3,naus,*knng,*pkg,*k,**l,*pk,*knn,n,**a;
	double xaus, x,*Ck;
	char filec[60];

	FILE *fp;

	fp=fopen("SC_d2_figure.edges","w");

	srand48(time(NULL));

	kgi=(int*)calloc(N,sizeof(int));
	kg=(int*)calloc(N,sizeof(int));
	k=(int*)calloc(N,sizeof(int));
	a=(int**)calloc(N,sizeof(int*));
	knng=(int*)calloc(N,sizeof(int));
	pkg=(int*)calloc(N,sizeof(int));
	knn=(int*)calloc(N,sizeof(int));
	Ck=(double*)calloc(N,sizeof(double));
	pk=(int*)calloc(N,sizeof(int));
	tri=(int***)calloc(N,sizeof(int**));

	for(i=0;i<N;i++){
		a[i]=(int*)calloc(N,sizeof(int));
        	tri[i]=(int**)calloc(2,sizeof(int*));
        	tri[i][0]=NULL;
		tri[i][1]=NULL;
        }

	xaus=4;
	while(xaus>3){
	/***********************************************************************************************/
	/* Initialization */
	for(i=0;i<N;i++){
	/* Nodes are assigned desired generalized degree according to a scale-free distribution */
		kgi[i]=(int)(m*pow(drand48(),-1./(gamma2-1.)));
		/* kgi[i]= poisson(lambda); */
		while(kgi[i]>(N-1)*(N-2)*0.5){ 
		/* Desired generalized degrees are re-drawn if they exceed the maximum possible generalized degree of a node (natural cut-off) */
			kgi[i]=(int)(m*pow(drand48(),-1./(gamma2-1.)));
			/* kgi[i]= poisson(lambda); */
		}
		kg[i]=0; /* Generalized degree of node i initially set to 0 */
		k[i]=0;  /* Degree of node i initially set to 0 */
		for(j=0;j<N;j++){
			a[i][j]=0;
		}

	}
		xaus=0;
		for(i=0;i<N;i++){
			xaus+=kgi[i];
		}

	naus=0; /* Back-track counter initially set to zero */
	/***********************************************************************************************/
	/* Stubs matched */
		while((xaus>3)&&(naus<1+Avoid*NX)){
			/*  Randomly select three nodes proportional to the number of unmatched stubs they have remaining. */

			x=xaus*drand48();
			i1=Choose(x);
			kg[i1]++;
			kgi[i1]--;
			xaus--;

			x=xaus*drand48();
			i2=Choose(x);
			kg[i2]++;
			kgi[i2]--;
			xaus--;

			x=xaus*drand48();
			i3=Choose(x);
			kg[i3]++;
			kgi[i3]--;
			xaus--;

			/* Check proposed matching is legal. */
			if((i1!=i2)&&(i2!=i3)&&(i3!=i1)&&(Check(i1,i2,i3)==0)){
                	/*Proposed matching legal. Create triangle and links.*/
				Triangle(i1,i2,i3);
				a[i1][i2]=1;
				a[i2][i1]=1;
				a[i1][i3]=1;
				a[i3][i2]=1;
				a[i2][i3]=1;
				a[i3][i1]=1;
			}
			else{
			/* Proposed matching illegal. Back-track and increment back-track counter by one. */
				naus++;
				if(Avoid==1){
					kg[i1]--;
					kgi[i1]++;
					kg[i2]--;
					kgi[i2]++;
					kg[i3]--;
					kgi[i3]++;
`				}
			}

		}
}

/*************************************************************************************************/
/* Degrees calculated */
	for (i=0;i<N;i++){
		for(j=i+1;j<N;j++){
			if(a[i][j]>0){
				k[i]++;
				k[j]++;
			}
		}
	}
/*************************************************************************************************/
/* Print list of edges to file */
	if (figure==1){
		for (i=0;i<N;i++){
			for(j=i+1;j<N;j++){
				if(a[i][j]==1){
					fprintf(fp,"%d %d\n",i,j);
				}
			}
		}
	}
/*************************************************************************************************/
	fclose(fp);

	return 0;
}



\end{lstlisting}


\subsubsection*{Code for $d=3$}
\begin{lstlisting}
/**************************************************************************************************
 * If you use this code, please cite G. Bianconi and O.T. Courtney
 * "Generalized network structures: the configuration model and the canonical ensemble of
 * simplicial complexes"
***************************************************************************************************
 * Code that  generates random simplicial complexes with scale-free generalized degree
 * distribution.
 *
 * The option to use a Poisson distributed generalized degree distribution has also been included.
 * The necessary code may be found in comments at the relevant points.
 *
 * This code uses:
 * N  Number of nodes in the simplicial complex
 * m  The minimum of the scale-free distribution
 * gamma2  Exponent of the scale-free distribution
 * lambda  Expected value of the Poisson distribution (commented-out)
 * Avoid  Whether or not 'back-tracking' is allowed when illegal matchings are proposed
 * (Avoid==1 allowed, Avoid==0 not allowed)
 * NX  Maximum number of 'back-tracks' before matching process restarts from an unmatched network
 *************************************************************************************************/

#include<stdio.h>
#include<stdlib.h>
#include<string.h>
#include<math.h>
#include<time.h>

#define N 10000
#define m 1
#define gamma2 2.8
/* #define lambda 10 */
#define Avoid 1
#define NX 80

int *kgi,*kg,***tri;

/*************************************************************************************************/
/* Randomly select an unmatched stub. Choose takes as its input a random number between 0 and the
total number of stubs and gives as its output the index of the node of the selected stub */
int Choose(double x){
	int i1,i;
	for (i=0;i<N;i++){
		x-=kgi[i];
		if (x<0){
			i1=i;
			break;
		}
	}
	return(i1);
}

/*************************************************************************************************/
/* Check if a tetrahedron exists. Takes four nodes as an input and outputs 1 if there already exists a triangle incident to them and 0 otherwise. */
int Check(i1,i2,i3,i4){
	int in,c=0;
	if ((i1==i2)||(i1==i3)||(i1==i4)||(i2==i3)||(i2==i4)||(i3==i4)){
		c=1;
	}
	if (c==0){
        for(in=0;in<kg[i1]-1;in++){
            if((tri[i1][0][in]==i2)&&(tri[i1][1][in]==i3)&&(tri[i1][2][in]=i4)){
                c=1;
                break;
            }
            if((tri[i1][0][in]==i2)&&(tri[i1][1][in]==i4)&&(tri[i1][2][in]=i3)){
                c=1;
                break;
            }
            if((tri[i1][0][in]==i3)&&(tri[i1][1][in]==i2)&&(tri[i1][2][in]=i4)){
                c=1;
                break;
            }
            if((tri[i1][0][in]==i3)&&(tri[i1][1][in]==i4)&&(tri[i1][2][in]=i2)){
                c=1;
                break;
            }
            if((tri[i1][0][in]==i4)&&(tri[i1][1][in]==i2)&&(tri[i1][2][in]=i3)){
                c=1;
                break;
            }
            if((tri[i1][0][in]==i4)&&(tri[i1][1][in]==i3)&&(tri[i1][2][in]=i2)){
                c=1;
                break;
            }
        }
	}
	return(c);
}

/*************************************************************************************************/
/* Create tetrahedron. Takes four nodes as an input and creates a tetrahedron incident to them. */
void Tetrahedron(int i1, int i2,int i3, int i4){
	int iaus;

	tri[i1][0]=(int*)realloc(tri[i1][0],kg[i1]*sizeof(int));
	tri[i1][1]=(int*)realloc(tri[i1][1],kg[i1]*sizeof(int));
	tri[i1][2]=(int*)realloc(tri[i1][2],kg[i1]*sizeof(int));

	tri[i2][0]=(int*)realloc(tri[i2][0],kg[i2]*sizeof(int));
	tri[i2][1]=(int*)realloc(tri[i2][1],kg[i2]*sizeof(int));
	tri[i2][2]=(int*)realloc(tri[i2][2],kg[i2]*sizeof(int));

	tri[i3][0]=(int*)realloc(tri[i3][0],kg[i3]*sizeof(int));
	tri[i3][1]=(int*)realloc(tri[i3][1],kg[i3]*sizeof(int));
	tri[i3][2]=(int*)realloc(tri[i3][2],kg[i3]*sizeof(int));

	tri[i4][0]=(int*)realloc(tri[i4][0],kg[i4]*sizeof(int));
	tri[i4][1]=(int*)realloc(tri[i4][1],kg[i4]*sizeof(int));
	tri[i4][2]=(int*)realloc(tri[i4][2],kg[i4]*sizeof(int));

	tri[i1][0][kg[i1]-1]=i2;
	tri[i1][1][kg[i1]-1]=i3;
	tri[i1][2][kg[i1]-1]=i4;

	tri[i2][0][kg[i2]-1]=i1;
	tri[i2][1][kg[i2]-1]=i3;
	tri[i2][2][kg[i2]-1]=i4;

	tri[i3][0][kg[i3]-1]=i1;
	tri[i3][1][kg[i3]-1]=i2;
	tri[i3][2][kg[i3]-1]=i4;

	tri[i4][0][kg[i4]-1]=i1;
	tri[i4][1][kg[i4]-1]=i2;
	tri[i4][2][kg[i4]-1]=i3;
}
/*************************************************************************************************/

int main(int argc, char** argv){
	int i,j,j2,i1,i2,i3,i4,naus,*knng,*pkg,*k,**l,*pk,*knn,n,**a,*Ck;
	double xaus, x;
	char filec[60];

	FILE *gp;

	srand48(time(NULL));

	gp=fopen("edge_list.txt","w");

	kgi=(int*)calloc(N,sizeof(int));
	kg=(int*)calloc(N,sizeof(int));
	k=(int*)calloc(N,sizeof(int));
	a=(int**)calloc(N,sizeof(int*));
	knng=(int*)calloc(N,sizeof(int));
	pkg=(int*)calloc(N,sizeof(int));
	knn=(int*)calloc(N,sizeof(int));
	pk=(int*)calloc(N,sizeof(int));
	Ck=(int*)calloc(N,sizeof(int));
	tri=(int***)calloc(N,sizeof(int**));
	for(i=0;i<N;i++){
            a[i]=(int*)calloc(N,sizeof(int));
            tri[i]=(int**)calloc(3,sizeof(int*));
            tri[i][0]=NULL;
            tri[i][1]=NULL;
            tri[i][2]=NULL;
    }

	xaus=4;
	while(xaus>3){
	/***********************************************************************************************/
	/* Initialization */
        for(i=0;i<N;i++){
        /* Nodes are assigned desired generalized degree according to a scale-free distribution */
            kgi[i]=(int)(m*pow(drand48(),-1./(gamma2-1.)));
            /* kgi[i]= poisson(lambda); */
            while(kgi[i]>pow(N,3.)/6){
            /* Desired generalized degrees are re-drawn if they exceed the maximum possible generalized degree of a node (natural cut-off) */
                kgi[i]=(int)(m*pow(drand48(),-1./(gamma2-1.)));
                /* kgi[i]= poisson(lambda); */
            }
            kg[i]=0;  /* Generalized degree of node iintially set to 0 */
            k[i]=0;   /* Degree of node i intially set to 0 */
            for(j=0;j<N;j++){
            a[i][j]=0;
            }

        }
        xaus=0;
        for(i=0;i<N;i++){
            xaus+=kgi[i];
        }
        naus=0; /* Back-track counter initially set to zero */
        /***********************************************************************************************/
        /* Stubs matched */
        while((xaus>4)&&(naus<1+Avoid*NX)){
        /* Randomly select four nodes proportional to the number of unmatched stubs they have remaining. */
            x=xaus*drand48();
            i1=Choose(x);
            kg[i1]++;
            kgi[i1]--;
            xaus--;

            x=xaus*drand48();
            i2=Choose(x);
            kg[i2]++;
            kgi[i2]--;
            xaus--;

            x=xaus*drand48();
            i3=Choose(x);
            kg[i3]++;
            kgi[i3]--;
            xaus--;

            x=xaus*drand48();
            i4=Choose(x);
            kg[i4]++;
            kgi[i4]--;
            xaus--;

            /* Check proposed matching is legal */
            if((Check(i1,i2,i3,i4)==0)){
            /* Proposed matching legal. Create tetrahedron and links */
                Tetrahedron(i1,i2,i3,i4);

                a[i1][i2]=1;
                a[i2][i1]=1;

                a[i1][i3]=1;
                a[i3][i1]=1;

                a[i1][i4]=1;
                a[i4][i1]=1;

                a[i2][i3]=1;
                a[i3][i2]=1;

                a[i2][i4]=1;
                a[i4][i2]=1;

                a[i3][i4]=1;
                a[i4][i3]=1;
            }
            else{
            /* Proposed matching illegal. Back-track and increment back-track counter by one */
                naus++;
                if(Avoid==1){
                    kg[i1]--;
                    kgi[i1]++;
                    kg[i2]--;
                    kgi[i2]++;
                    kg[i3]--;
                    kgi[i3]++;
                    kg[i4]--;
                    kgi[i4]++;
                }
            }

        }
    }
/*************************************************************************************************/
/* Degrees calculated */
	for (i=0;i<N;i++){
		for(j=i+1;j<N;j++){
			if(a[i][j]>0){
				k[i]++;
				k[j]++;
			}
		}
	}
/*************************************************************************************************/
/* Print list of edges to file */
	for (i=0;i<N;i++){
		for(j=0;j<N;j++){
			if(a[i][j]==1){
				fprintf(gp,'%d %d\n',i,j)
			}
		}
	}
/*************************************************************************************************/
	fclose(gp);

	return 0;
}

\end{lstlisting}

\end{document}